\documentclass[pra,12pt,tightenlines]{revtex4}
\usepackage{latexsym}
\usepackage{graphicx}
\usepackage{amsmath}
\usepackage{amsbsy}
\usepackage{amsthm}
\usepackage{bbm}
\usepackage{bm}
\newtheorem{thm}{Theorem}
\newtheorem{lem}{Lemma}

\begin{document}

\title{Bloch vectors for qudits and geometry of entanglement}
\author{Reinhold A. Bertlmann} \email{reinhold.bertlmann@univie.ac.at}
\author{Philipp Krammer} \email{philipp.krammer@univie.ac.at}
\affiliation{Faculty of Physics, University of Vienna, Boltzmanngasse 5, A-1090 Vienna,
Austria}

\begin{abstract}

We present three different matrix bases that can be used to decompose density matrices of
$d$--dimensional quantum systems, so-called qudits: the \emph{generalized Gell-Mann
matrix basis}, the \emph{polarization operator basis}, and the \emph{Weyl operator
basis}. Such a decomposition can be identified with a vector ---the Bloch vector, i.e. a
generalization of the well known qubit case--- and is a convenient expression for
comparison with measurable quantities and for explicit calculations avoiding the handling
of large matrices. We consider the important case of an isotropic two--qudit state and
decompose it according to each basis. Investigating the geometry of entanglement of
special parameterized two--qubit and two--qutrit states, in particular we calculate the
Hilbert--Schmidt measure of entanglement, we find that the Weyl operator basis is the
optimal choice since it is closely connected to the entanglement of the considered
states.

\end{abstract}

\maketitle

\section{Introduction}

The state of a $d$--dimensional quantum system ---a qudit--- is usually described by a $d
\times d$ density matrix. For high dimensions, where the matrices become large (for
composite systems of $n$ particles the matrices are of even much larger dimension
$d^{\,n} \times d^{\,n}$), a simple way to express density matrices is of great interest.

Since the space of matrices is a vector space, there exist bases of matrices which can be
used to decompose any matrix. For qubits such a basis contains the three Pauli matrices,
accordingly, a density matrix can be expressed by a 3--dimensional vector, the
\emph{Bloch vector} and any such vector has to lie within the so-called \emph{Bloch ball}
\cite{bloch46, nielsen00}. Unique for qubits is the fact that any point on the sphere,
\emph{Bloch sphere}, and inside the ball corresponds to a physical state, i.e. a density
matrix. The pure states lie on the sphere and the mixed ones inside.

In higher dimensions there exist different matrix bases that can be used to express
qudits as ($d^{\,2}-1$)--dimensional vectors as well. Different to the qubit case,
however, is that the map induced is not bijective: not every point on the ``Bloch
sphere'' in dimensions $d^{\,2}-1$ corresponds to a physical state. Nevertheless the
vectors are often also called ``Bloch vectors'' (see in this context, e.g.,
Refs.~\cite{kimura03, kryszewski06, jakobczyk01, kimura05, mendas06}).

In this paper we want to present and compare three different matrix bases for a Bloch
vector decomposition of qudits. In Sec.~\ref{prelim} we propose the properties of any
matrix basis for using it as a ``practical'' decomposition of density matrices and recall
the general notation of Bloch vectors. In Secs.~\ref{secggb} -- \ref{secwob} we offer
three different matrix bases: the \emph{generalized Gell-Mann matrix basis}, the
\emph{polarization operator basis}, and the \emph{Weyl operator basis}. For all these
bases we give examples in the dimensions of our interest and present the different Bloch
vector decompositions of an arbitrary density matrix in the standard matrix notation.
Next in Sec.~\ref{seciso}, by constructing tensor products of states we study the
isotropic two--qudit state and present the results for the three matrix decompositions,
i.e. for the three different Bloch vectors. In Sec.~\ref{secapp} we focus on the geometry
of entanglement of the isotropic two--qudit state and calculate the Hilbert--Schmidt
measure of entanglement (see, e.g., Refs.~\cite{witte99, ozawa00, bertlmann02,
bertlmann05}). Its connection to the optimal entanglement witness is shown, which is
determined in terms of the three matrix bases. Furthermore, we calculate explicitly the
Hilbert--Schmidt measure for so-called two--parameter states, which define a plane in the
Hilbert--Schmidt space and provide a picture for the geometry of entanglement of states
in higher dimensions, in particular we studied the two--qubit (4--dimensions) and
two--qutrit states (9--dimensions). The mathematical and physical
advantages/disadvantages by using the three different matrix bases are discussed in
Sec.~\ref{conclusion}, where also the final conclusions are drawn.

\section{Preliminaries}
\label{prelim}

A \emph{qudit} state is represented by a density operator in the Hilbert--Schmidt space acting on
the d--dimensional Hilbert space ${\cal H}^{\,d}$ that can be written as a matrix ---the density
matrix--- in the \emph{standard basis} $ \left\{ \left| k
\right\rangle \right\}, \,$ with $k=1,2, \ldots d \,$ or $k=0,1,2, \ldots d-1$.\\

\emph{Properties of a ``practical'' matrix basis.} For practical reasons the general properties of
a matrix basis which is used for the Bloch vector decomposition of qudits are the following:
\begin{enumerate}
    \item[i)] The basis includes the identity matrix $\mathbbm{1}$ and $d-1$ matrices
    $\left\{ A_i \right\}$ of dimension $d \times d$ which are traceless, i.e. $\textnormal{Tr}
    A_i = 0 \,$.
    \item[ii)] The matrices of any basis $\left\{ A_i \right\}$ are orthogonal, i.e.
    \begin{equation} \label{orthogon}
    \textnormal{Tr}\, A_i^{\dag} A_j \,=\, N \, \delta_{ij} \quad \textrm{with} \quad N\in
    \mathbbm{R}\,.\\
    \end{equation}
\end{enumerate}

\emph{Bloch vector expansion of a density matrix.} Since any matrix in the
Hilbert-Schmidt space of dimension $d$ can be decomposed with a matrix basis $\left\{ A_i
\right\}$, we can of course decompose a qudit density matrix as well and get the
\emph{Bloch vector expansion} of the density matrix,
\begin{equation} \label{bvgenbasis}
    \rho \;=\; \frac{1}{d} \,\mathbbm{1} \,+\, \vec{b} \cdot \vec{\Gamma} \,,
\end{equation}
where $\vec{b} \cdot\vec{\Gamma}$ is a linear combination of all matrices $\left\{ A_i \right\}$
and the vector $\vec{b} \in \mathbbm{R}^{d^2-1}$ with $b_i = \langle \Gamma_i \rangle =
\textnormal{Tr}\rho\Gamma_i\,$ is called \emph{Bloch vector}. The term
$\frac{1}{d} \mathbbm{1}$ is fixed because of condition $\text{Tr} \rho = 1$.\\

\emph{Remark.} Note that a given density matrix $\rho$ can always be decomposed into a
Bloch vector, but not any vector $\sigma$ that is of the form \eqref{bvgenbasis} is
automatically a density matrix, even if it satisfies the conditions Tr$\sigma = 1$ and
Tr$\sigma^2 \leq 1$ since generally it does not imply $\sigma \geq 0$.

Each different matrix basis induces a different Bloch vector lying within a Bloch
hypersphere where, however, not every point of the hypersphere corresponds to a physical
state (with $\rho \geq 0$); these points are excluded (holes). The geometric character of
the Bloch space in higher dimensions turns out to be quite complicated and is still of
great interest (see Refs.~\cite{kimura03, kryszewski06, jakobczyk01, kimura05,
mendas06}).

All different Bloch hyperballs are isomorphic since they correspond to the same density
matrix $\rho$. The interesting question is which Bloch hyperball ---which matrix basis---
is optimal for a specific purpose, like the calculation of the entanglement degree or the
determination of the geometry of the Hilbert space or the comparison with measurable
quantities.

\section{The Generalized Gell-Mann Matrix Basis}
\label{secggb}

\subsection{Definition and example}

The generalized Gell-Mann matrices (GGM) are higher--dimensional extensions of the Pauli
matrices (for qubits) and the Gell-Mann matrices (for qutrits), they are the standard
SU(N) generators (in our case $N=d$). They are defined as three different types of
matrices and for simplicity we use here the operator notation; then the density matrices
follow by simply writing the operators in the standard basis (see, e.g.
Refs.~\cite{kimura03, rakotonirina05}):
\begin{enumerate}
\item[i)]{$\frac{d(d-1)}{2}$ symmetric GGM}
    \begin{equation} \label{ggms}
        \Lambda^{jk}_s \;=\; | j \rangle \langle k | \,+\, | k \rangle
        \langle j |\,, \quad 1 \leq j < k \leq d \;,
    \end{equation}
\item[ii)]{$\frac{d(d-1)}{2}$ antisymmetric GGM}
    \begin{equation} \label{ggma}
        \Lambda^{jk}_a \;=\; -i \,| j \rangle \langle k | \,+\, i \,| k \rangle
        \langle j |\,, \quad 1 \leq j < k \leq d \;,
    \end{equation}
\item[iii)]{$(d-1)$ diagonal GGM}
    \begin{equation} \label{ggmd}
        \Lambda^l \;=\; \sqrt{\frac{2}{l(l+1)}} \left( \sum_{j=1}^{l} |
        j \rangle \langle j | \,-\, l \,| l+1 \rangle \langle l+1 |
        \right), \quad 1 \leq l \leq d-1 \;.
    \end{equation}
\end{enumerate}
In total we have $d^2-1$ GGM; it follows from the definitions that all GGM are Hermitian
and traceless. They are orthogonal and form a basis, the generalized Gell-Mann matrix
Basis (GGB). A proof for the orthogonality of GGB we present in the
Appendix \ref{orthogonality}.\\

\emph{Examples.} Let us recall the case of dimension $3$, the $8$ Gell-Mann matrices (for
a representation see, e.g., Refs.~\cite{bertlmann05,caves00})
\begin{enumerate}
\item[i)]{$3$ symmetric Gell-Mann matrices}
\begin{eqnarray} \label{defgellmann-s}
    & \lambda^{12}_s = \left(
    \begin{array}{ccc}
        0 & 1 & 0 \\
        1 & 0 & 0 \\
        0 & 0 & 0 \\
    \end{array} \right), \quad
    \lambda^{13}_s = \left(
    \begin{array}{ccc}
        0 & 0 & 1 \\
        0 & 0 & 0 \\
        1 & 0 & 0 \\
    \end{array} \right), \quad
    \lambda^{23}_s = \left(
    \begin{array}{ccc}
        0 & 0 & 0 \\
        0 & 0 & 1 \\
        0 & 1 & 0 \\
    \end{array} \right), &
    \nonumber \\
\end{eqnarray}
\item[ii)]{$3$ antisymmetric Gell-Mann matrices}
\begin{eqnarray} \label{defgellmann-a}
    & \lambda^{12}_a = \left(
    \begin{array}{ccc}
        0 & -i & 0 \\
        i & 0 & 0 \\
        0 & 0 & 0 \\
    \end{array} \right), \quad
    \lambda^{13}_a = \left(
    \begin{array}{ccc}
        0 & 0 & -i \\
        0 & 0 & 0 \\
        i & 0 & 0 \\
    \end{array} \right), \quad
    \lambda^{23}_a = \left(
    \begin{array}{ccc}
        0 & 0 & 0 \\
        0 & 0 & -i \\
        0 & i & 0 \\
    \end{array} \right), &
    \nonumber \\
\end{eqnarray}
\item[ii)]{$2$ diagonal Gell-Mann matrices}
\begin{eqnarray} \label{defgellmann-d}
    & \lambda^1 = \left(
    \begin{array}{ccc}
        1 & 0 & 0 \\
        0 & -1 & 0 \\
        0 & 0 & 0 \\
    \end{array} \right), \quad
    \lambda^2 = \frac{1}{\sqrt{3}} \left(
    \begin{array}{ccc}
        1 & 0 & 0 \\
        0 & 1 & 0 \\
        0 & 0 & -2 \\
    \end{array} \right). &
\end{eqnarray}
\end{enumerate}

To see how they generalize for higher dimensions we show the case we need for qudits of
dimension $d=4\,$:
\begin{enumerate}
\item[i)]{$6$ symmetric GGM}
\begin{eqnarray} \label{ggm4s}
    & \Lambda^{12}_s = \left(
    \begin{array}{cccc}
        0 & 1 & 0 & 0 \\
        1 & 0 & 0 & 0 \\
        0 & 0 & 0 & 0 \\
        0 & 0 & 0 & 0 \\
    \end{array} \right), \quad
    \Lambda^{13}_s = \left(
    \begin{array}{cccc}
        0 & 0 & 1 & 0 \\
        0 & 0 & 0 & 0 \\
        1 & 0 & 0 & 0 \\
        0 & 0 & 0 & 0 \\
    \end{array} \right), \quad
    \Lambda^{14}_s = \left(
    \begin{array}{cccc}
        0 & 0 & 0 & 1 \\
        0 & 0 & 0 & 0 \\
        0 & 0 & 0 & 0 \\
        1 & 0 & 0 & 0 \\
    \end{array} \right), &
    \nonumber \\
    & \Lambda^{23}_s = \left(
    \begin{array}{cccc}
        0 & 0 & 0 & 0 \\
        0 & 0 & 1 & 0\\
        0 & 1 & 0 & 0 \\
        0 & 0 & 0 & 0 \\
    \end{array} \right), \quad
    \Lambda^{24}_s = \left(
    \begin{array}{cccc}
        0 & 0 & 0 & 0 \\
        0 & 0 & 0 & 1 \\
        0 & 0 & 0 & 0 \\
        0 & 1 & 0 & 0 \\
    \end{array} \right), \quad
    \Lambda^{34}_s = \left(
    \begin{array}{cccc}
        0 & 0 & 0 & 0 \\
        0 & 0 & 0 & 0 \\
        0 & 0 & 0 & 1 \\
        0 & 0 & 1 & 0 \\
    \end{array} \right), &
\end{eqnarray}
\item[ii)]{$6$ antisymmetric GGM}
\begin{eqnarray} \label{ggm4a}
    & \Lambda^{12}_a = \left(
    \begin{array}{cccc}
        0 & -i & 0 & 0 \\
        i & 0 & 0 & 0 \\
        0 & 0 & 0 & 0 \\
        0 & 0 & 0 & 0 \\
    \end{array} \right), \quad
    \Lambda^{13}_a = \left(
    \begin{array}{cccc}
        0 & 0 & -i & 0 \\
        0 & 0 & 0 & 0 \\
        i & 0 & 0 & 0 \\
        0 & 0 & 0 & 0 \\
    \end{array} \right), \quad
    \Lambda^{14}_a = \left(
    \begin{array}{cccc}
        0 & 0 & 0 & -i \\
        0 & 0 & 0 & 0 \\
        0 & 0 & 0 & 0 \\
        i & 0 & 0 & 0 \\
    \end{array} \right), &
    \nonumber \\
    & \Lambda^{23}_a = \left(
    \begin{array}{cccc}
        0 & 0 & 0 & 0 \\
        0 & 0 & -i & 0\\
        0 & i & 0 & 0 \\
        0 & 0 & 0 & 0 \\
    \end{array} \right), \quad
    \Lambda^{24}_a = \left(
    \begin{array}{cccc}
        0 & 0 & 0 & 0 \\
        0 & 0 & 0 & -i \\
        0 & 0 & 0 & 0 \\
        0 & i & 0 & 0 \\
    \end{array} \right), \quad
    \Lambda^{34}_a = \left(
    \begin{array}{cccc}
        0 & 0 & 0 & 0 \\
        0 & 0 & 0 & 0 \\
        0 & 0 & 0 & -i \\
        0 & 0 & i & 0 \\
    \end{array} \right), &
\end{eqnarray}
\item[iii)]{$3$ diagonal GGM}
\begin{equation} \label{ggm4d}
    \Lambda^{1} = \left(
    \begin{array}{cccc}
        1 & 0 & 0 & 0 \\
        0 & -1 & 0 & 0 \\
        0 & 0 & 0 & 0 \\
        0 & 0 & 0 & 0 \\
    \end{array} \right), \quad
    \Lambda^{2} = \frac{1}{\sqrt{3}} \left(
    \begin{array}{cccc}
        1 & 0 & 0 & 0 \\
        0 & 1 & 0 & 0 \\
        0 & 0 & -2 & 0 \\
        0 & 0 & 0 & 0 \\
    \end{array} \right), \quad
    \Lambda^{3} = \frac{1}{\sqrt{6}} \left(
    \begin{array}{cccc}
        1 & 0 & 0 & 0 \\
        0 & 1 & 0 & 0 \\
        0 & 0 & 1 & 0 \\
        0 & 0 & 0 & -3 \\
    \end{array} \right).
\end{equation}
\end{enumerate}

Using the GGB we obtain, in general, the following Bloch vector expansion of a density
matrix
\begin{equation} \label{bvggb-d}
    \rho \;=\; \frac{1}{d} \,\mathbbm{1} \,+\, \vec{b} \cdot \vec{\Lambda} \,,
\end{equation}
with the Bloch vector $\vec{b} = \big(\{b^{jk}_s\},\{b^{jk}_a\},\{b^l\}\big)\,$, where
the components are ordered and for the indices we have the restrictions $1 \leq j < k
\leq d$ and $1 \leq l \leq d-1\,$. The components are given by $b^{jk}_s =
\textrm{Tr}\Lambda^{jk}_s\rho\,$, $b^{jk}_a = \textrm{Tr}\Lambda^{jk}_a\rho$ and $b^l =
\textrm{Tr}\Lambda^l\rho\,$. All Bloch vectors lie within a hypersphere of radius
$|\vec{b}| \leq \sqrt{(d-1)/2d}\,$. For example, for qutrits the Bloch vector components
are $\vec{b} = \big(b^{12}_s,b^{13}_s,b^{23}_s,b^{12}_a,b^{13}_a,b^{23}_a,b^1,b^2\big)$
corresponding to the Gell-Mann matrices \eqref{defgellmann-s}, \eqref{defgellmann-a},
\eqref{defgellmann-d} and $|\vec{b}| \leq \sqrt{1/3}\,$.

As already mentioned the allowed range of $\vec{b}$ is restricted. It has an interesting
geometric structure which has been calculated analytically for the case of qutrits by
studying $2$--dimensional planes in the $8$--dimensional Bloch space \cite{kimura03} or
numerically by considering $3$--dimensional cross--sections \cite{mendas06}. In any case,
pure states lie on the surface and the mixed ones inside.

\subsection{Standard matrix basis expansion by GGB}

The standard matrices are simply the $d \times d$ matrices that have only one entry $1$
and the other entries $0$ and form an orthonormal basis of the Hilbert--Schmidt space. We
write these matrices shortly as operators
\begin{equation} \label{standardmatrices}
    | j \rangle \langle k | \,, \qquad \textrm{with} \quad j,k = 1, \ldots, d \,.
\end{equation}
Any matrix can easily be decomposed into a ``vector'' via a certain linear combination of
the matrices \eqref{standardmatrices}. Knowing the expansion of matrices
\eqref{standardmatrices} into GGB we can therefore find the decomposition of any matrix
in terms of the GGB.

We find the following expansion of standard matrices
\eqref{standardmatrices} into GGB \,:
\begin{equation} \label{smggb}
    | j \rangle \langle k | \;=\; \begin{cases}
        \frac{1}{2} \left( \Lambda^{jk}_s \,+\, i \Lambda^{jk}_a \right)
            & \text{for } j < k \\
        \frac{1}{2} \left( \Lambda^{kj}_s \,-\, i \Lambda^{kj}_a \right)
            & \text{for } j > k \\
        - \sqrt{\frac{j-1}{2j}} \,\Lambda^{j-1} \,+\, \sum\limits_{n=0}^{d-j-1}
            \frac{1}{\sqrt{2(j+n)(j+n+1)}} \,\Lambda^{j+n} \,+\, \frac{1}{d}
            \,\mathbbm{1} & \text{for } j = k \,.
        \end{cases}
\end{equation}

\emph{Proof.} The first two cases can be easily verified.

To show the last case we first set up a recurrence relation for $| l \rangle \langle l|$, which we
obtain by eliminating the term $\sum_{j=1}^{l-1} | j \rangle \langle j|$ in the two expressions
\eqref{ggmd} for $\Lambda^l$ and $\Lambda^{l-1}\,$
\begin{equation} \label{recl}
    | l \rangle \langle l| \;=\; - \sqrt{\frac{l-1}{2l}} \,\Lambda^{l-1} \,+\,
    \sqrt{\frac{l+1}{2l}} \,\Lambda^l \,+\, |l+1\rangle \langle l+1| \,,
\end{equation}
and we consider the case $l+1=d$
\begin{equation} \label{recd}
| d-1 \rangle \langle d-1| \;=\; - \sqrt{\frac{d-2}{2(d-1)}} \,\Lambda^{d-2} \,+\,
    \sqrt{\frac{d}{2(d-1)}} \,\Lambda^{d-1} \,+\, |d\rangle \langle d| \,.
\end{equation}
From $\Lambda^{d-1}$ given by Eq.~\eqref{ggmd}
\begin{equation}
    \Lambda^{d-1} \;=\; \sqrt{\frac{2}{(d-1)d}} \,\left( \sum_{j=1}^{d-1}
    |j \rangle \langle j | \,-\, (d-1) |d \rangle \langle d| \right) \,,
\end{equation}
we get the Bloch vector decomposition of $|d\rangle \langle d|$
\begin{equation}\label{d-state-decomp}
    |d \rangle \langle d| \;=\; \frac{1}{d} \left(
    -\sqrt{\frac{(d-1)d}{2}} \,\Lambda^{d-1} \,+\, \mathbbm{1} \right) \,,
\end{equation}
where we have applied $\sum_{j=1}^{d-1} |j \rangle \langle j | \,=\, \mathbbm{1} - |d
\rangle
    \langle d|\,$.

Inserting now decomposition \eqref{d-state-decomp} into relation \eqref{recd} we gain the
Bloch vector expansion for $|d-1 \rangle \langle d-1|$ and recurrence relation
\eqref{recl} provides $|d-2 \rangle \langle d-2|$ and so forth. Thus finally we find
\begin{equation} \label{dminusn}
    |d-n \rangle \langle d-n | \;=\; - \sqrt{\frac{d-n-1}{2(d-n)}}
    \,\Lambda^{d-n-1} \,+\, \sum_{k=0}^{n-1} \frac{1}{\sqrt{2(d-n+k+1)(d-n+k)}}
    \,\Lambda^{d-n+k} + \frac{1}{d} \mathbbm{1} ,
\end{equation}
the relation we had to prove, where $d-n=j\,$. $\Box$

\section{The polarization operator basis}

\subsection{Definition and examples}

\emph{Definition.} The polarization operators in the Hilbert-Schmidt
space of dimension $d$ are defined as the following $d \times d$
matrices \cite{varshalovich88, kryszewski06} \,:
\begin{equation} \label{defpo}
    T_{LM} \;=\; \sqrt{\frac{2L+1}{2s+1}} \sum_{k,l =1}^d C^{s m_k}_{s
    m_l , \,LM} \,|k \rangle \langle l | \,.
\end{equation}
The used indices have the properties
\begin{eqnarray}
    & s = \frac{d-1}{2} \,, & \nonumber\\
    & L = 0,1, \ \ldots \ ,2s \,, & \nonumber\\
    & M = -L, -L+1, \ldots, L-1, L \,, & \nonumber\\
    & m_1 = s, \ m_2 = s-1, \ldots ,m_d = -s \,.
\end{eqnarray}
The coefficients $C^{s m_k}_{s m_l , \,LM}$ are identified with the usual Clebsch--Gordan
coefficients $C^{j m}_{j_1 m_1 , \,j_2 m_2}$ of the angular momentum theory and are
displayed explicitly in tables, e.g., in Ref.~\cite{varshalovich88}.\\

For $L=M=0$ the polarization operator is proportional to the identity matrix
\cite{varshalovich88, kryszewski06},
\begin{equation} \label{po00}
    T_{00} \,=\, \frac{1}{\sqrt{d}} \, \mathbbm{1} \,.
\end{equation}
It is shown in Ref.~\cite{kryszewski06} that all polarization operators (except $T_{00}$)
are traceless, in general \emph{not} Hermitian), and that orthogonality relation
\eqref{orthogon} is satisfied
\begin{equation}\label{orthogonal-pob}
    \text{Tr} \, T_{L_1 M_1}^\dag T_{L_2 M_2} \;=\; \delta_{L_1 L_2}
    \delta_{M_1 M_2} \;.
\end{equation}
Therefore the $d^2$ polarization operators \eqref{defpo} form an orthonormal matrix basis
---the polarization operator basis (POB)--- of the Hilbert--Schmidt space of dimension
$d$.\\

\emph{Examples.} The simplest example is of dimension $2$, the qubit. For a qubit the POB
is given by the following matrices ($s = 1/2; L=0,1; M=-1,0,1$)
\begin{eqnarray}\label{pobdim2}
    & T_{00} \;=\; \frac{1}{\sqrt{2}} \left(
    \begin{array}{cc}
        1 & 0 \\
        0 & 1
    \end{array} \right), \quad
    T_{11} \;=\; - \left(
    \begin{array}{cc}
        0 & 1 \\
        0 & 0
    \end{array} \right), & \nonumber\\
    & T_{10} \;=\; \frac{1}{\sqrt{2}} \left(
    \begin{array}{cc}
        1 & 0 \\
        0 & -1
    \end{array} \right), \quad
    T_{1-1} \;=\; \left(
    \begin{array}{cc}
        0 & 0 \\
        1 & 0
    \end{array} \right). &
\end{eqnarray}

For the next higher dimension $d=3$ ($s=1$), the case of qutrits, we
get $9$ polarization operators $T_{LM}$ with $L=0,1,2$ and
$M=-L,...,L\,$. According to definition \eqref{defpo} they are as
follows ($T_{00}=\frac{1}{\sqrt{3}}\,\mathbbm{1}$) \,:
\begin{eqnarray} \label{pobdim3}
    & T_{11} = -\frac{1}{\sqrt{2}}\left(
    \begin{array}{ccc}
        0 & 1 & 0 \\
        0 & 0 & 1 \\
        0 & 0 & 0 \\
    \end{array} \right), \quad
    T_{10} = \frac{1}{\sqrt{2}}\left(
    \begin{array}{ccc}
        1 & 0 & 0 \\
        0 & 0 & 0 \\
        0 & 0 & -1 \\
    \end{array} \right), \quad
    T_{1-1} = \frac{1}{\sqrt{2}}\left(
    \begin{array}{ccc}
        0 & 0 & 0 \\
        1 & 0 & 0 \\
        0 & 1 & 0 \\
    \end{array} \right), &
    \nonumber \\
    & T_{22} = \left(
    \begin{array}{ccc}
        0 & 0 & 1 \\
        0 & 0 & 0 \\
        0 & 0 & 0 \\
    \end{array} \right), \quad
    T_{21} = \frac{1}{\sqrt{2}}\left(
    \begin{array}{ccc}
        0 & -1 & 0 \\
        0 & 0 & 1 \\
        0 & 0 & 0 \\
    \end{array} \right), \quad
    T_{20} = \frac{1}{\sqrt{6}}\left(
    \begin{array}{ccc}
        1 & 0 & 0 \\
        0 & -2 & 0 \\
        0 & 0 & 1 \\
    \end{array} \right), &
    \nonumber \\
    &
    T_{2-1} = \frac{1}{\sqrt{2}}\left(
    \begin{array}{ccc}
        0 & 0 & 0 \\
        1 & 0 & 0 \\
        0 & -1 & 0 \\
    \end{array} \right), \quad
    T_{2-2} = \left(
    \begin{array}{ccc}
        0 & 0 & 0 \\
        0 & 0 & 0 \\
        1 & 0 & 0 \\
    \end{array} \right). &
\end{eqnarray}

Then the decomposition of any density matrix into a Bloch vector by
using the POB has, in general, the following form \,:
\begin{equation} \label{bvpob-d}
    \rho \;=\; \frac{1}{d} \,\mathbbm{1} \,+\, \sum_{L=1}^{2s} \sum_{M=-L}^L b_{LM} T_{LM}
     \;=\; \frac{1}{d} \,\mathbbm{1} \,+\, \vec{b} \cdot\vec{T} \,,
\end{equation}
with the Bloch vector $\vec{b} =
(b_{1-1},b_{10},b_{11},b_{2-2},b_{2-1},b_{20},...,b_{LM})$, where the components are
ordered and given by $b_{LM} = \textnormal{Tr}\,T_{LM}^\dagger \rho\,$. In general the
components $b_{LM}$ are complex since the polarization operators $T_{LM}$ are not
Hermitian. All Bloch vectors lie within a hypersphere of radius $|\vec{b}| \leq
\sqrt{(d-1)/d}\,$.

In $2$ dimensions the Bloch vector $\vec{b} = (b_{1-1},b_{10},b_{11})$ is limited by
$|\vec{b}| \leq \frac{1}{\sqrt{2}}\,$ and forms a spheroid \cite{kryszewski06}, the pure
states occupy the surface and the mixed ones lie in the volume. This decomposition is
fully equivalent to the standard description of Bloch vectors with Pauli matrices.

In higher dimensions, however, the structure of the allowed range of $\vec{b}$ (due to
the positivity requirement $\rho \geq 0$) is quite complicated, as can be seen already
for $d=3$ (for details see Ref.~\cite{kryszewski06}). Nevertheless, pure states are on
the surface, mixed ones lie within the volume and the maximal mixed one corresponds to
$|\vec{b}|=0\,$, thus $|\vec{b}|$ is a kind of measure for the mixedness of a quantum
state.

\subsection{Standard matrix basis expansion by POB}

The standard matrices \eqref{standardmatrices} can be expanded by the POB as follows
\cite{varshalovich88}
\begin{equation} \label{smpob}
| i \rangle \langle j| \;=\; \sum_L \sum_M \sqrt{\frac{2L+1}{2s+1}} \, C^{s m_i}_{s
m_j,\,LM} \,T_{LM} \,.
\end{equation}
Note that $\sum_M$ is actually fixed by the condition $m_j + M = m_i$.\\

\emph{Proof}. Inserting definition \eqref{defpo} on the right--hand side (RHS) of
equation \eqref{smpob} we find
\begin{eqnarray}
    \textrm{RHS} & \;=\; & \sum_{k,l} \left( \sum_L \frac{2L+1}{2s+1}
        \,C^{sm_i}_{sm_j,\,LM} \,C^{sm_k}_{sm_l,\,LM} \right) |k \rangle
        \langle l| = \nonumber\\
    & \;=\; & \sum_{k,l} \delta_{jl} \,\delta_{ik} \,|k \rangle \langle
        l| \nonumber\\
    & \;=\; & |i \rangle \langle j| \,,
\end{eqnarray}
where we used the sum rule for Clebsch--Gordan coefficients \cite{varshalovich88}
\begin{equation}
    \sum_{c,\gamma} \frac{2c+1}{2b+1} \,C^{b \beta}_{a \alpha,\,c \gamma}
    \,C^{b \beta'}_{a \alpha',\,c \gamma} \;=\; \delta_{\alpha \alpha'}
    \,\delta_{\beta \beta'} \,.
\end{equation}

\section{Weyl operator basis} \label{secwob}

\subsection{Definition and example} \label{secwob-defandexample}

Finally we want to discuss a basis of the Hilbert--Schmidt space of dimension $d$ that
consists of the following $d^2$ operators
\begin{equation} \label{defwo}
    U_{nm} \;=\; \sum_{k=0}^{d-1} e^{\frac{2 \pi i}{d}\,kn} \,| k \rangle
    \langle (k+m) \,\textrm{mod}\,d| \qquad n,m = 0,1, \ldots ,d-1 \,,
\end{equation}
where we use the standard basis of the Hilbert space.

These operators have been introduced in the context of quantum teleportation of qudit states
\cite{bennett93} and are often called \emph{Weyl operators} in the literature (see e.g.
Refs.~\cite{narnhofer06, baumgartner06}). The $d^2$ operators \eqref{defwo} are unitary and form
an orthonormal basis of the Hilbert--Schmidt space (a proof is presented in
Appendix~\ref{secproofon}) -- the Weyl operator basis (WOB). They can be used to create a basis of
$d^2$ maximally entangled qudit states \cite{narnhofer06, werner01, vollbrecht00}.

Clearly the operator $U_{00}$ represents the identity $U_{00} = \mathbbm{1}\,$.\\

\emph{Example.} Let us show the example of dimension $3$, the qutrit case. There the Weyl
operators \eqref{defwo} have the following matrix form
\begin{eqnarray}
    & 
    U_{01} = \begin{pmatrix}
        0 & 1 & 0 \\
        0 & 0 & 1 \\
        1 & 0 & 0
        \end{pmatrix}, \qquad\qquad\quad
    U_{02} = \begin{pmatrix}
        0 & 0 & 1 \\
        1 & 0 & 0 \\
        0 & 1 & 0
        \end{pmatrix}, & \\
    & U_{10} = \begin{pmatrix}
        1 & 0 & 0 \\
        0 & e^{2 \pi i/3} & 0 \\
        0 & 0 & e^{-2 \pi i/3}
        \end{pmatrix}, \quad
    U_{11} = \begin{pmatrix}
        0 & 1 & 0 \\
        0 & 0 & e^{2 \pi i/3} \\
        e^{-2 \pi i/3} & 0 & 0
        \end{pmatrix}, \quad
    U_{12} = \begin{pmatrix}
        0 & 0 & 1 \\
        e^{2 \pi i/3} & 0 & 0 \\
        0 & e^{-2 \pi i/3} & 0
        \end{pmatrix}, & \nonumber\\
    & U_{20} = \begin{pmatrix}
        1 & 0 & 0 \\
        0 & e^{-2 \pi i/3} & 0 \\
        0 & 0 & e^{2 \pi i/3}
        \end{pmatrix}, \quad
    U_{21} = \begin{pmatrix}
        0 & 1 & 0 \\
        0 & 0 & e^{-2 \pi i/3} \\
        e^{2 \pi i/3} & 0 & 0
        \end{pmatrix}, \quad
    U_{22} = \begin{pmatrix}
        0 & 0 & 1 \\
        e^{-2 \pi i/3} & 0 & 0 \\
        0 & e^{2 \pi i/3} & 0
        \end{pmatrix}. & \nonumber
\end{eqnarray}\\

Using the WOB we can decompose quite generally any density matrix into a Bloch vector
\begin{equation} \label{bvwob-d}
    \rho \;=\; \frac{1}{d} \,\mathbbm{1} \,+\, \sum_{n,m=0}^{d-1} b_{nm} U_{nm}
     \;=\; \frac{1}{d} \,\mathbbm{1} \,+\, \vec{b} \cdot \vec{U} \,,
\end{equation}
with $n,m = 0,1, ... ,d-1$ ($b_{00}=0$). The components of the Bloch vector $\vec{b} =
\big(\{b_{nm}\}\big)$ are ordered and given by $b_{nm} =
\textnormal{Tr}\,U_{nm}\,\rho\,$. In general the components $b_{nm}$ are complex since
the Weyl operators are not Hermitian and the complex conjugates fulfil the relation
$b_{n\,m}^\ast = e^{\frac{2 \pi i}{d}\,nm} \, b_{{-n}{-m}}\,$, which follows easily from
definition \eqref{defwo} together with the hermiticity of $\rho\,$.

All Bloch vectors lie within a hypersphere of radius $|\vec{b}| \leq \sqrt{d-1}/d\,$. For
example, for qutrits the Bloch vector is expressed by
$\vec{b}=(b_{01},b_{02},b_{10},b_{11},b_{12},b_{20},b_{21},b_{22}$) and $|\vec{b}| \leq
\sqrt{2}/3\,$. In $3$ and higher dimensions the allowed range of the Bloch vector is
quite restricted within the hypersphere and the detailed structure is not known yet.

Note that in $2$ dimensions the WOB as well as the GGB coincides with the Pauli matrix
basis and the POB represents a rotated Pauli basis (where $\sigma_{\pm} =
\frac{1}{2}\,(\sigma_1 \pm i\sigma_2)$), in particular
\begin{eqnarray}
\left\{U_{00},U_{01},U_{10},U_{11}\right\} \;&=&\;
\left\{\mathbbm{1},\sigma_1 ,\sigma_3 , i\sigma_2 \right\}
\,,\\
\left\{\mathbbm{1},\lambda^{12}_s,\lambda^{12}_a,\lambda^1 \right\} \;&=&\;
\left\{\mathbbm{1},\sigma_1 ,\sigma_2 , \sigma_3 \right\} \,,\\
\left\{T_{00},T_{11},T_{10},T_{1-1}\right\} \;&=&\;
\left\{\frac{1}{\sqrt{2}}\,\mathbbm{1},\,-\sigma_{+}
,\,\frac{1}{\sqrt{2}}\,\sigma_3 , \,\sigma_{-} \right\} \,.
\end{eqnarray}

\subsection{Standard matrix basis expansion by WOB}

The standard matrices \eqref{standardmatrices} can be expressed by the WOB in the
following way
\begin{equation} \label{smwob}
    |j \rangle \langle k | \;=\; \frac{1}{d}\, \sum_{l=0}^{d-1} e^{-\frac{2 \pi i}{d}
    \,lj} \,U_{l\, (k-j)\,\textrm{mod}\,d} \;.
\end{equation}

\emph{Proof.} We insert the definition of the Weyl operators \eqref{defwo} on the
right--hand side (RHS) of Eq.~\eqref{smwob}, use Eq.~\eqref{complexsumrule} and get
\begin{eqnarray}
    \textrm{RHS} & \;=\; & \frac{1}{d} \,\sum_{l,r=0}^{d-1} e^{\frac{2 \pi i}{d}\,l(r-j)}
    \,|r \rangle \langle (r+k-j)\,\textrm{mod}\,d| \nonumber\\
    & \;=\; & |j \rangle \langle k | \;+\; \frac{1}{d} \sum_{r \neq j,\,r=0}^{d-1}
        \sum_{l=0}^{d-1} e^{\frac{2 \pi i}{d}\,l(r-j)} \,|r \rangle \langle
        (r+k-j)\,\textrm{mod}\,d | \nonumber\\
    & \;=\; & |j \rangle \langle k | \,. \quad\Box
\end{eqnarray}

\section{Isotropic two--qudit state}
\label{seciso}

Now we consider bipartite systems in a $d\times d$ dimensional Hilbert space ${\cal
H}^{\,d}_A \otimes {\cal H}^{\,d}_B$. The observables acting in the subsystems ${\cal
H}_A$ and ${\cal H}_B$ are usually called Alice and Bob in quantum communication.

Quite generally, a density matrix of a two--qudit state acting on ${\cal H}^{\,d}_A
\otimes {\cal H}^{\,d}_B$ can be decomposed in the following way (neglecting the
reference to $A$ and $B$)
\begin{equation} \label{2-quditstategeneral}
    \rho \;=\; \frac{1}{d} \,\mathbbm{1} \otimes \mathbbm{1} \,+\,
    n_i\,\Gamma_i \otimes \mathbbm{1} \,+\, m_i\,\mathbbm{1} \otimes \Gamma_i \,+\,
    c_{ij}\,\Gamma_i \otimes \Gamma_j \,, \qquad
    n_i, m_i, c_{ij} \in \mathbbm{C} \,,
\end{equation}
where $\left\{ \Gamma_i \right\}$ represents some basis in the subspace ${\cal
H}^{\,d}\,$. The term $c_{ij}\,\Gamma_i \otimes \Gamma_j$ always can be diagonalized by
two independent orthogonal transformations on $\Gamma_i$ and $\Gamma_j$ \cite{henley62}.
Altogether there are $(d^{\,2})^2 - 1$ independent terms.

However, for isotropic two--qudit states ---the case we consider in our paper--- the
second and third term in expression \eqref{2-quditstategeneral} vanish and the fourth
term reduces to $c_{ii}\,\Gamma_i \otimes \Gamma_i$, which implies the vanishing of
$(d^{\,2} - 1)^2 + (d^{\,2} -1) = d^{\,2}(d^{\,2} - 1)$ terms. Consequently, for an
isotropic two--qudit density matrix there remain $d^{\,2} - 1$ independent terms, which
provides the dimension of the corresponding Bloch vector. Thus the isotropic two--qudit
Bloch vector is of the same dimension ---lives in the same subspace--- as the one--qudit
vector, which is a comfortable simplification.

Explicitly, the \emph{isotropic} two--qudit state $\rho_{\alpha}^{(d)}$ is defined as
follows \cite{horodecki99, rains99, horodecki01} \,:
\begin{equation} \label{rhodefiso}
    \rho_{\alpha}^{(d)} \;=\; \alpha \left| \phi_+^d \right\rangle \left\langle \phi_+^d \right|
    \,+\, \frac{1-\alpha}{d^2}\,\mathbbm{1}\,, \quad \alpha \in \mathbbm{R}\,, \quad
    - \frac{1}{d^2-1} \leq \alpha \leq 1 \;,
\end{equation}
where the range of $\alpha$ is determined by the positivity of the state. The state
$\left| \phi^d_+ \right\rangle$, a Bell state, is maximally entangled and given by
\begin{equation} \label{defmaxent}
    \left| \phi^d_+ \right\rangle \;=\; \frac{1}{\sqrt{d}} \,
    \sum_j \left| j \right\rangle \otimes \left| j \right\rangle\;,
\end{equation}
where $\left\{ \left| j \right\rangle \right\}$ denotes the standard basis of the
d--dimensional Hilbert space.

\subsection{Expansion into GGB}

Let us first calculate the Bloch vector notation for the Bell state $\left| \phi_+^d
\right\rangle \left\langle \phi_+^d \right|$ in the GGB. It is convenient to split the
state into two parts
\begin{eqnarray}\label{maxentstandbasis}
    \left| \phi_+^d \right\rangle \left\langle \phi_+^d
        \right| & \;=\; & \frac{1}{d} \sum_{j,k =1}^{d} |j \rangle \langle k |
        \otimes |j \rangle \langle k| \nonumber\\
    & \;=\; & A \,+\, B \,,
\end{eqnarray}
where $A$ and $B$ are defined by
\begin{eqnarray}\label{isoggmA}
    A \;&:=&\; \frac{1}{d} \sum_{j < k} |j \rangle \langle k| \otimes | j
    \rangle \langle k | \,+\, \frac{1}{d} \sum_{j < k} |k \rangle \langle j| \otimes | k
    \rangle \langle j | \,,\\
    B \;&:=&\; \frac{1}{d} \sum_{j} |j \rangle \langle j| \otimes |
    j \rangle \langle j | \,,\label{isoggmB}
\end{eqnarray}
and to calculate the two terms separately.

For term $A$ we use the standard matrix expansion \eqref{smggb} for the case $j \neq k$
and get
\begin{eqnarray}
    A & \;=\; & \frac{1}{4d} \left[ \sum_{j<k} \left( \Lambda^{jk}_s + i
        \Lambda^{jk}_a \right) \otimes \left( \Lambda^{jk}_s + i
        \Lambda^{jk}_a \right) \,+\, \sum_{j<k} \left( \Lambda^{jk}_s - i
        \Lambda^{jk}_a \right) \otimes \left( \Lambda^{jk}_s - i
        \Lambda^{jk}_a \right) \right] \nonumber\\
    & \;=\; & \frac{1}{2d} \,\sum_{i<j} \left( \Lambda^{jk}_s \otimes
    \Lambda^{jk}_s \,-\, \Lambda^{jk}_a \otimes
    \Lambda^{jk}_a \right) \,.
\end{eqnarray}
For term $B$ we need the case $j = k$ in expansion \eqref{smggb} and obtain after some
calculations (the details are presented in Appendix \ref{termB})
\begin{eqnarray}\label{ggb-term-B}
    B \;\;=\;\; \frac{1}{2d} \,
    \sum_{m=1}^{d-1} \Lambda^m \otimes \Lambda^m \,+\, \frac{1}{d^2} \,\mathbbm{1} \otimes
    \mathbbm{1} \,.
\end{eqnarray}
Thus all together we find the following GGB Bloch vector notations, for the Bell state
\eqref{maxentstandbasis}
\begin{eqnarray}\label{maxentggb}
    \left| \phi_+^d \right\rangle \left\langle \phi_+^d \right|  \;\;=\;\;
    \frac{1}{d^2} \,\mathbbm{1} \otimes \mathbbm{1} \,+\, \frac{1}{2d} \;\Lambda \,,
\end{eqnarray}
and for the isotropic two--qudit state \eqref{rhodefiso}
\begin{eqnarray}\label{isoggb}
    \rho^{(d)}_\alpha \;\;=\;\; \frac{1}{d^2} \,\mathbbm{1} \otimes
    \mathbbm{1} \,+\, \frac{\alpha}{2d} \;\Lambda \,,
\end{eqnarray}
where we defined
\begin{eqnarray}\label{Bloch-Lambda}
    \Lambda \;:=\;\; \sum_{i<j} \Lambda^{jk}_s \otimes
    \Lambda^{jk}_s \,-\,  \sum_{i<j} \Lambda^{jk}_a \otimes
    \Lambda^{jk}_a \,+\, \sum_{m=1}^{d-1} \Lambda^m \otimes \Lambda^m \,.
\end{eqnarray}

\subsection{Expansion into POB}

Now we calculate the Bell state $\left| \phi_+^d \right\rangle \left\langle \phi_+^d
\right|$ in the POB. Using expansion \eqref{smpob} and the sum rule for the
Clebsch--Gordan coefficients \cite{varshalovich88}
\begin{equation}\label{CGsumrule}
    \sum_{\alpha, \gamma} C^{c \gamma}_{a \alpha , b \beta} \,C^{c \gamma}_{a \alpha ,
    b' \beta '} \;=\; \frac{2c+1}{2b+1} \,\,\delta_{b b'} \,\delta_{\beta \beta '} \,,
\end{equation}
we obtain
\begin{eqnarray} \label{maxentpob}
    \left| \phi_+^d \right\rangle \left\langle \phi_+^d
        \right| & \;=\; & \frac{1}{d} \sum_{i,j =1}^{d} |i \rangle \langle j |
        \otimes |i \rangle \langle j| \nonumber\\
    & \;=\; & \frac{1}{d} \sum_{L,L'} \frac{\sqrt{(2L+1)(2L'+1)}}{2s+1} \left(
        \sum_{i,j} C^{s m_i}_{s m_j , LM} C^{s m_i}_{s m_j, L' M}
        \right) T_{LM} \otimes T_{L'M} \nonumber\\
    & \;=\; & \frac{1}{d} \sum_{L,L'} \frac{\sqrt{(2L+1)(2L'+1)}}{2L+1}
        \,\, \delta_{L,L'} \, T_{LM} \otimes T_{L' M} \nonumber\\
    & \;=\; & \frac{1}{d} \sum_{L} T_{LM} \otimes T_{LM} \nonumber\\
    & \;=\; & \frac{1}{d^2} \,\mathbbm{1} \otimes \mathbbm{1} \,+\,
    \frac{1}{d} \, T \,,
\end{eqnarray}
where we extracted the unity (recall Eq.~\eqref{po00}) and defined
\begin{equation}\label{Bloch-T}
    T \;:=\; \sum_{L,M \neq 0,0} T_{LM} \otimes T_{LM} \,.
\end{equation}
Result \eqref{maxentpob} provides the POB Bloch vector notation of the isotropic
two--qudit state \eqref{rhodefiso}
\begin{equation}\label{isopob}
    \rho^{(d)}_\alpha \;=\; \frac{1}{d^2} \,\mathbbm{1} \otimes
    \mathbbm{1} \,+\, \frac{\alpha}{d} \, T \,.
\end{equation}

\subsection{Expansion into WOB}

Finally we present the Bell state in the WOB (the details for our approach using the
standard matrix expression \eqref{smwob} can be found in the Appendix \ref{BellWOB}, see
also Ref.~\cite{narnhofer06})
\begin{equation} \label{maxentwob2}
    \left| \phi_+^d \right\rangle \left\langle \phi_+^d \right| \;=\;
    \frac{1}{d^2} \,\mathbbm{1} \otimes \mathbbm{1} \,+\, \frac{1}{d^2} \, U \,,
\end{equation}
with
\begin{equation} \label{defu}
    U \;:=\; \sum_{l,m = 0}^{d-1} U_{lm} \otimes
        U_{-lm} \,, \qquad (l,m) \neq (0,0) \,,
\end{equation}
where negative values of the index $l$ have to be considered as $mod \ d\,$, and from
formula \eqref{maxentwob2} we find the WOB Bloch vector notation of the isotropic
two--qudit state
\begin{equation} \label{isowob}
    \rho^{(d)}_\alpha \;=\; \frac{1}{d^2} \,\mathbbm{1} \otimes \mathbbm{1}
    \,+\, \frac{\alpha}{d^2} \, U \,.
\end{equation}

\section{Hilbert--Schmidt measure --- Applications of the matrix bases}\label{secapp}

\subsection{Entangled isotropic two--qudit states}\label{secqudits}

In Ref.~\cite{bertlmann05} the connection between the Hilbert--Schmidt (HS) measure of
entanglement \cite{witte99,ozawa00,bertlmann02} and the optimal entanglement witness is
investigated. Explicit calculations for both quantities are presented in case of
isotropic qutrit states. For higher dimensions, the isotropic two--qudit states, the
above quantities are determined as well but in terms of a rather general matrix basis
decomposition. With the results of the present paper we can calculate all quantities
explicitly. Let us recall the basic notations we need.

The HS \emph{measure} is defined as the minimal HS distance of an
entangled state $\rho_{\rm{ent}}$ to the set of separable states $S$
\begin{equation} \label{defhs}
    D(\rho_{\rm{ent}}) \;:=\; \min_{\rho \in S} \left\| \rho - \rho_{\rm{ent}} \right\|
    \;=\; \left\| \rho_0 - \rho_{\rm{ent}} \right\| \,,
\end{equation}
where $\rho_0$ denotes the nearest separable state, the minimum of the HS distance.

\begin{figure}
    \centering
        \includegraphics[width=0.40\textwidth]{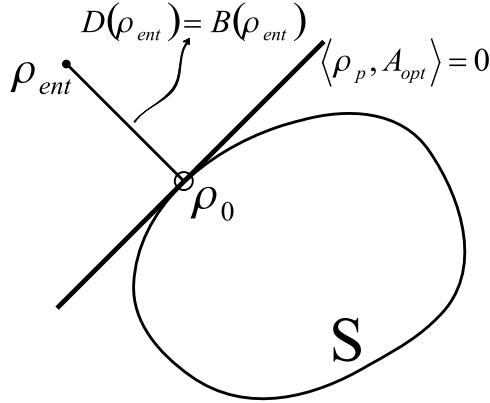}
    \caption{Illustration of the Bertlmann--Narnhofer--Thirring Theorem~\eqref{bnttheorem}}
    \label{figbnttheo}
\end{figure}

An \emph{entanglement witness} $A \in {\cal A}$ (${\cal A} = {\cal A}_A \otimes {\cal
A}_B\,$, the HS space of operators acting on the Hilbert space of states) is a Hermitian
operator that ``detects'' the entanglement of a state $\rho_{\rm ent}$ via inequalities
\cite{horodecki96, terhal00, terhal02, bertlmann02}
\begin{eqnarray} \label{defentwit}
    \left\langle \rho_{\rm ent},A \right\rangle \;=\; \textnormal{Tr}\, \rho_{\rm ent} A
    & \;<\; & 0 \,,\nonumber\\
    \left\langle \rho,A \right\rangle = \textnormal{Tr}\, \rho A & \;\geq\; & 0 \qquad
    \forall \rho \in S \,.
\end{eqnarray}
An entanglement witness is ``optimal'', denoted by $A_{\rm{opt}}\,$, if apart from
Eq.~(\ref{defentwit}) there exists a separable state $\rho_0 \in S$ such that
\begin{equation}
    \left\langle \rho_0 ,A_{\rm{opt}} \right\rangle \;=\; 0 \,.
\end{equation}
The operator $A_{\rm{opt}}$ defines a tangent plane to the set of separable states $S$
and all states $\rho_p$ with $\left\langle \rho_p ,A_{\rm{opt}} \right\rangle \;=\; 0 \,$
lie within that plane; see Fig.~\ref{figbnttheo}.
\\

According to Ref.~\cite{bertlmann02}, we call the lower one of the inequalities
(\ref{defentwit}) a \textit{generalized Bell inequality}, short GBI. ``Generalized''
means that it detects entanglement and not just non--locality. Re-writing
Eq.~\eqref{defentwit} as
\begin{equation} \label{gbi}
    \left\langle \rho,A \right\rangle \,-\, \left\langle \rho_{\rm ent},A \right\rangle
    \;\geq\; 0 \qquad \forall \rho \in S \,,
\end{equation}
the \emph{maximal violation} of the GBI is defined by
\begin{equation} \label{maxviolationgbi}
    B(\rho_{\rm ent}) \;=\; \max_{A, \, \left\| A - a \mathbbm{1} \right\| \leq 1}
    \left( \min_{\rho \in S} \left\langle \rho,A \right\rangle \,-\,
    \left\langle \rho_{\rm ent},A \right\rangle \right),
\end{equation}
where the maximum is taken over all possible entanglement witnesses $A$, suitably
normalized.\\

Then an interesting connection between the HS measure and the concept of entanglement
witnesses is given by the Bertlmann--Narnhofer--Thirring Theorem, illustrated in
Fig.~\ref{figbnttheo} \cite{bertlmann02}.
\begin{thm}\label{BNT-theorem}
\hspace{2cm}
\begin{enumerate}
    \item[i)] The maximal violation of the GBI is equal to the minimal distance of
    $\rho_{\rm ent}$ to the set $S$
    \begin{equation} \label{bnttheorem}
    D(\rho_{\rm ent}) \;=\; B(\rho_{\rm ent}) \;.
    \end{equation}
    \item[ii)] The maximal violation of the GBI is attained for an optimal entanglement
    witness
    \begin{equation} \label{entwitmaxviolation}
    A_{opt} \;=\; \frac{\rho_0 - \rho_{\rm ent} \,-\, \left\langle \rho_0 ,
    \rho_0 - \rho_{\rm ent} \right\rangle \mathbbm{1}}{\left\| \rho_0 - \rho_{\rm ent}
    \right\|}\;.
    \end{equation}
\end{enumerate}
\end{thm}
Thus the calculation of the optimal entanglement witness $A_{\rm{opt}}$ to a given
entangled state $\rho_{\rm ent}$ reduces to the determination of the nearest separable
state $\rho_0\,$. In special cases $\rho_0$ might be easy to find but in general its
detection is quite a difficult task.\\

Now let us apply the matrix bases we discussed in the previous sections and calculate the
quantities introduced above. As an entangled state we consider the isotropic two--qudit
state $\rho_\alpha^{(d), \, \rm{ent}}$, that is the state $\rho_{\alpha}^{(d)}$
\eqref{rhodefiso} for $\frac{1}{d+1} < \alpha \leq 1$.

Starting with the GGB we can express that state in our Bloch vector notation by formula
\eqref{isoggb}. Then the nearest separable state is reached at $\alpha = \frac{1}{d+1}$
\begin{equation} \label{quditisoresult1}
    \rho^{(d)}_0 \;=\; \rho_{\alpha = \frac{1}{d+1}}^{(d)} \;=\;
    \frac{1}{d^2} \,\mathbbm{1} \otimes \mathbbm{1} \,+\,
    \frac{1}{2 \, d(d+1)} \, \Lambda \,.
\end{equation}
It provides the HS measure
\begin{equation} \label{quditisoresult2}
    D(\rho_{\alpha , \,\rm{ent}}^{(d)}) \;=\; \left\| \rho_0^{(d)} -
        \rho_{\alpha , \,\rm{ent}}^{(d)} \right\| \;=\; \frac{\sqrt{d^2-1}}{d}\,
        \left( \alpha \,-\, \frac{1}{d+1} \right) \,,
\end{equation}
and the optimal entanglement witness \eqref{entwitmaxviolation}
\begin{equation} \label{quditisoresult3}
    A_{\rm{opt}}(\rho_{\alpha , \,\rm{ent}}^{(d)}) \;=\; \frac{1}{d} \,\sqrt{\frac{d-1}{d+1}}\,
    \mathbbm{1} \otimes \mathbbm{1} \,-\, \frac{1}{2\sqrt{d^2-1}} \; \Lambda \,,
\end{equation}
where we used the HS norm $\| \Lambda \| = 2 \sqrt{d^2-1}\,$.

Clearly, the maximal violation of the GBI $B$ equals the HS measure $D$
\begin{eqnarray}\label{BequalsD}
    B(\rho_{\alpha , \,\rm{ent}}^{(d)}) &\;=\;& - \left\langle \rho_{\alpha , \,\rm{ent}}^{(d)} ,
    A_{\rm{opt}} \right\rangle \nonumber\\
    &\;=\;& \frac{\sqrt{d^2-1}}{d} \left( \alpha \,-\, \frac{1}{d+1} \right) \;=\;
    D(\rho_{\alpha , \,\rm{ent}}^{(d)}) \,.
\end{eqnarray}

For expressing above quantities by the matrix bases POB and WOB it suffices to calculate
the proportionality factors between $\Lambda$, $T$ and $U\,$. By comparison of the three
forms for the isotropic qudit state \eqref{isoggb}, \eqref{isopob} and \eqref{isowob} we
find
\begin{equation}
    \Lambda \;=\; 2 \, T \qquad  \textrm{and} \qquad T \;=\; \frac{1}{d} \, U \,.
\end{equation}
It provides the following expressions, for the POB
\begin{equation}
    \rho^{(d)}_0 \;=\; \rho_{\alpha = \frac{1}{d+1}}^{(d)} \;=\;
    \frac{1}{d^2} \,\mathbbm{1} \otimes \mathbbm{1} \,+\,
    \frac{1}{d(d+1)} \; T \,,
\end{equation}
\begin{equation}
    A_{\rm{opt}}(\rho_{\alpha , \,\rm{ent}}^{(d)}) \;=\; \frac{1}{d} \,\sqrt{\frac{d-1}{d+1}}
    \,\mathbbm{1} \otimes \mathbbm{1} \,-\, \frac{1}{\sqrt{d^2-1}} \; T \,,
\end{equation}
and for the WOB
\begin{equation}
    \rho^{(d)}_0 \;=\; \rho_{\alpha = \frac{1}{d+1}}^{(d)} \;=\;
    \frac{1}{d^2} \,\mathbbm{1} \otimes \mathbbm{1} \,+\,
    \frac{1}{d^2(d+1)} \; U \,,
\end{equation}
\begin{equation}
    A_{\rm{opt}}(\rho_{\alpha , \,\rm{ent}}^{(d)}) \;=\; \frac{1}{d} \,\sqrt{\frac{d-1}{d+1}}
    \,\mathbbm{1} \otimes \mathbbm{1} \,-\, \frac{1}{d\sqrt{d^2-1}} \; U \,.
\end{equation}
Of course, the HS measure $D(\rho_{\alpha , \,\rm{ent}}^{(d)})$ remains the same
expression \eqref{quditisoresult2} independent of the chosen matrix basis, which can
easily be verified using $\| T \| = \sqrt{d^2-1}$ and $\| U \| = d \sqrt{d^2-1}\,$.

\subsection{Two--parameter entangled states --- qubits}\label{secqubits}

\begin{figure}
  \begin{centering}
    \includegraphics[width=0.7\textwidth]{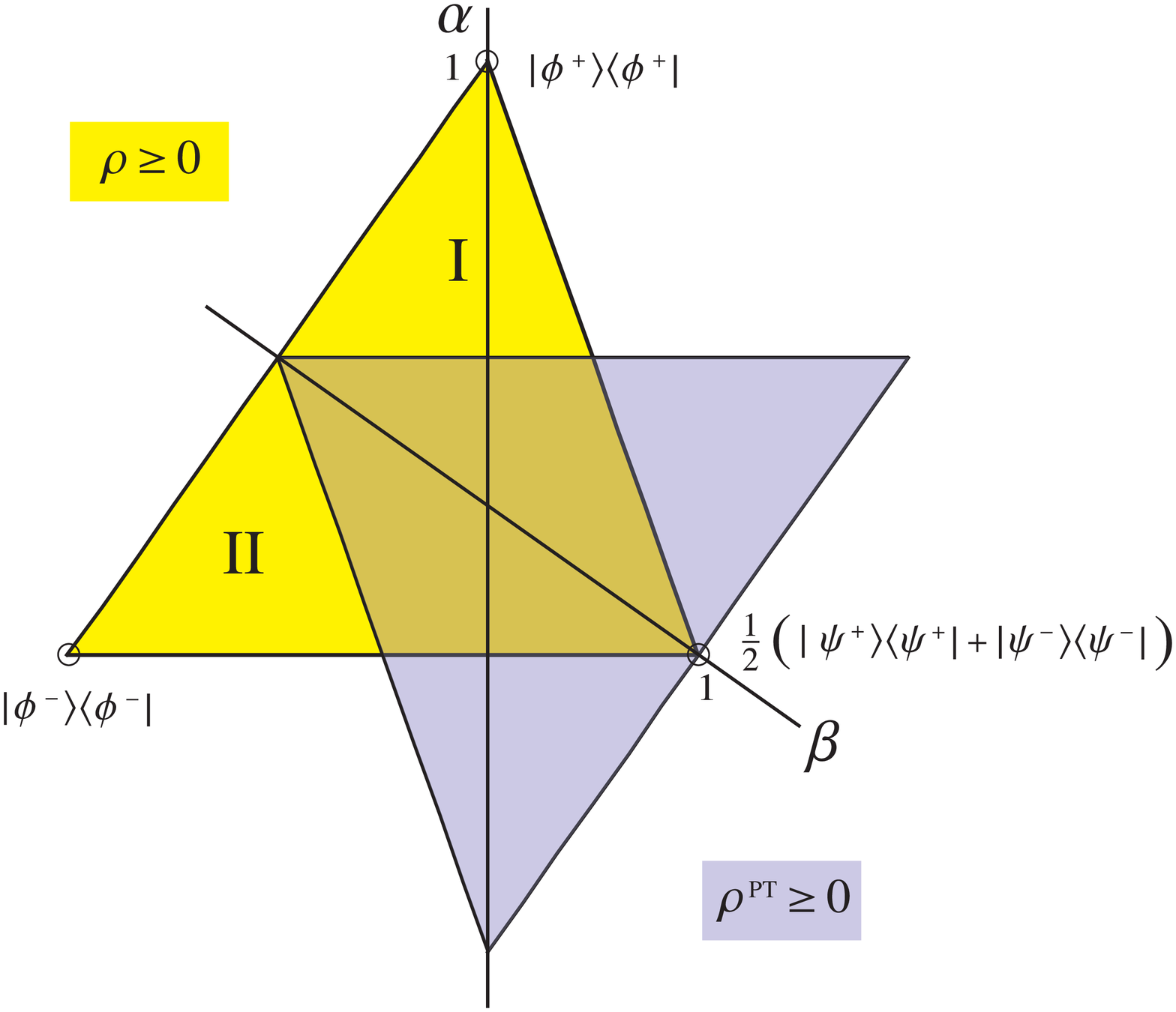}\\
    \caption{Illustration of the two-qubit states $\rho_{\alpha, \beta}$ \eqref{rhoqubit}.}
    \label{figqubit}
  \end{centering}
\end{figure}

As an application of our Bloch vector notation in the several matrix bases we want to
determine the HS measure of entanglement for the following two--qubit states which are a
particular mixture of the Bell states $| \psi^- \rangle, | \psi^+ \rangle, | \phi^-
\rangle, | \phi^+ \rangle$
\begin{equation} \label{rhoqubit}
    \rho_{\alpha, \beta} \;=\; \frac{1-\alpha-\beta}{4} \,\mathbbm{1} \,+\,
    \alpha | \phi^+ \rangle \langle \phi^+ | \,+\, \frac{\beta}{2}
    \left( | \psi^+ \rangle \langle \psi^+ | \,+\, | \psi^- \rangle
    \langle \psi^- | \right) \,.
\end{equation}
The states \eqref{rhoqubit} are characterized by the two parameters $\alpha$ and $\beta$
and we will refer to the states as the \emph{two--parameter states}. Of course, the
positivity requirement $ \rho_{\alpha, \beta} \geq 0$ constrains the possible values of
$\alpha$ and $\beta$, namely
\begin{equation}
    \alpha \;\leq\; - \beta +1, \quad\; \alpha \;\geq\; \frac{1}{3} \beta - \frac{1}{3},
    \quad\; \alpha \;\leq\; \beta+1 \,,
\end{equation}
which geometrically corresponds to a triangle, see Fig.~\ref{figqubit}.

According to Peres \cite{peres96} and the Horodeckis \cite{horodecki96} the separability
of the states is determined by the \emph{positive partial transposition criterion} (PPT),
at least in dimensions $2 \otimes 2$ and $2 \otimes 3\,$. States \eqref{rhoqubit} which
are positive under partial transposition have the following constraints
\begin{equation}
    \alpha \;\geq\; \beta -1, \quad\; \alpha \;\leq\; \frac{1}{3} \beta + \frac{1}{3},
    \quad\; \alpha \geq -\beta-1 \,,
\end{equation}
and correspond to the rotated triangle; then the overlap, a rhombus, represents the
separable states, see Fig.~\ref{figqubit}.

In the picture drawn in Fig.~\ref{figqubit} the orthogonal lines are indeed orthogonal in
HS space. Therefore the coordinate axes for the parameter $\alpha$ and $\beta$ are
necessarily non--orthogonal. In particular, the $\alpha$ axis has to be orthogonal to the
boundary line $\alpha = - \beta-1$, and the $\beta$ axis has to be orthogonal to $\alpha
= \beta+1$.

\begin{figure}
  \begin{centering}
    \includegraphics[width=0.5\textwidth]{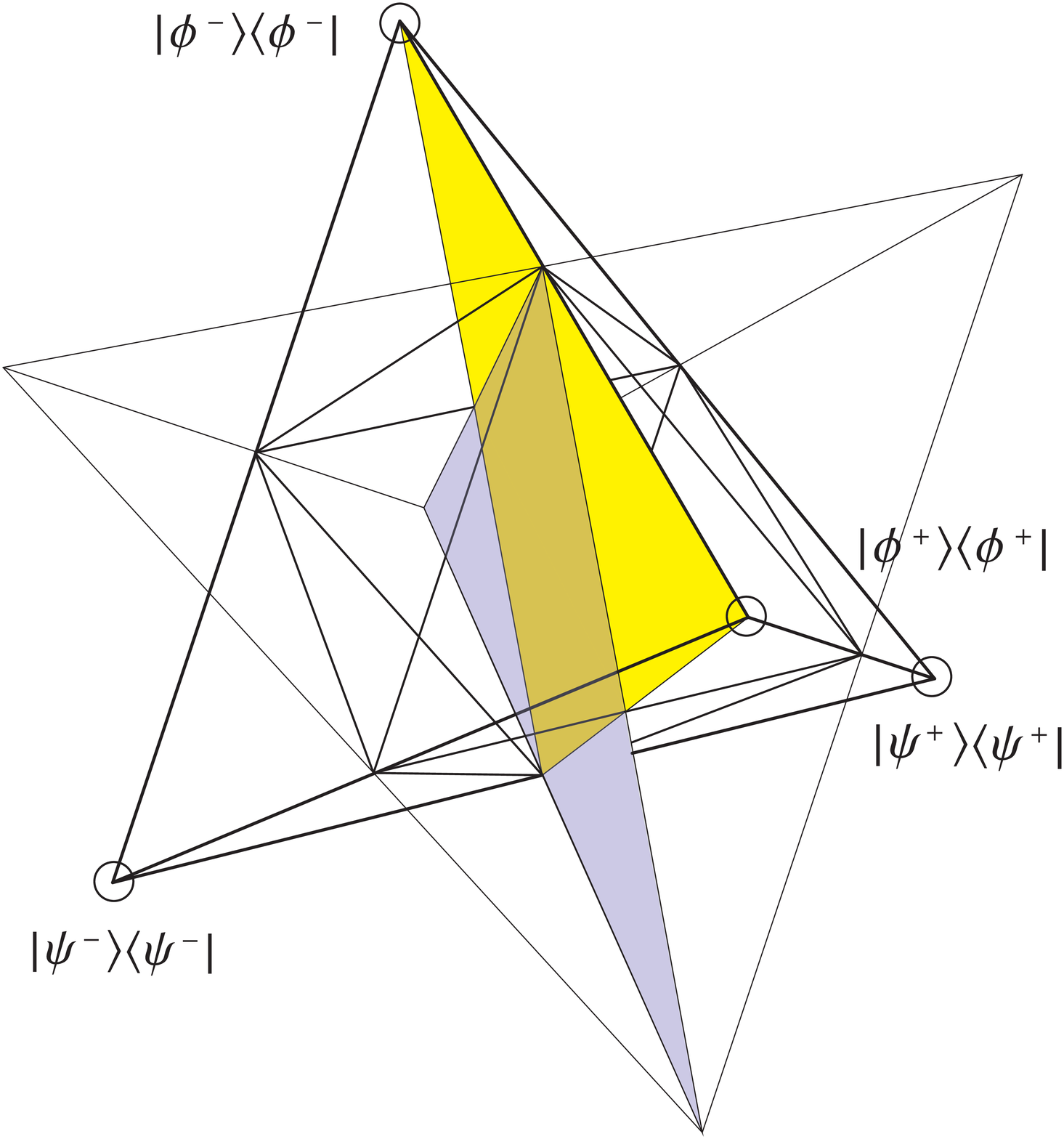}\\
    \caption{Location of the plane of states $\rho_{\alpha, \beta}$
    \eqref{rhoqubit} in the tetrahedron formed by the Bell states.}
    \label{figtetrahedron}
  \end{centering}
\end{figure}

The two--parameter states $\rho_{\alpha, \beta}$ define a plane in the HS space. It is
quite illustrative to see how this plane is located in the 3--dimensional spin space of
the density matrices, where the 4 Bell states form a tetrahedron due to the positivity
condition \cite{bertlmann02, vollbrecht01, horodecki96b}. Applying PPT the tetrahedron is
rotated producing an intersection ---a double pyramid--- which represents the separable
states. This is shown in Fig.~\ref{figtetrahedron}.

To calculate the HS measure \eqref{defhs} for the two--parameter qubit state
\eqref{rhoqubit} we express the state in terms of the Pauli matrix basis, which is indeed
just the GGB and equivalent to the WOB for dimension $d=2$ (see
Sec.~\ref{secwob-defandexample})
\begin{equation} \label{rhoqubitpauli}
    \rho_{\alpha,\beta} \;=\; \frac{1}{4} \left( \mathbbm{1} \,+\, \alpha
    \left( \sigma_1 \otimes \sigma_1 \,-\, \sigma_2 \otimes \sigma_2
    \right) \,+\, \left( \alpha - \beta \right)
    \sigma_3 \otimes \sigma_3 \right) \,,
\end{equation}
where we have used the well--known Pauli matrix decomposition of the Bell states (see, e.g.,
Ref.~\cite{bertlmann02}).\\

In order to determine the HS measure for the entangled two--parameter states
$\rho_{\alpha,\beta}^{\rm{ent}}$ we have to find the nearest separable states, which is
usually the most difficult task to perform in this context. In Ref.~\cite{bertlmann05} a
lemma is presented to check if a particular separable state is indeed the nearest
separable state to a given entangled one:
\begin{lem} \label{lemsepablenearest}
    A state $\tilde{\rho}$ is equal to the nearest separable state $\rho_0$
    if and only if the operator
    \begin{equation} \label{ctilde}
    \tilde{C} \;=\; \frac{\tilde{\rho} - \rho_{\rm ent} \,-\, \left\langle \tilde{\rho} ,
    \tilde{\rho} - \rho_{\rm ent} \right\rangle \mathbbm{1}}{\left\| \tilde{\rho} -
    \rho_{\rm ent} \right\|}
\end{equation}
is an entanglement witness.
\end{lem}

Lemma~\ref{lemsepablenearest} is used here in the following way. First, we calculate the
separable state that has the nearest Euclidean distance in the geometric picture
(Fig.~\ref{figqubit}) and call this state $\tilde{\rho}$. But since the regarded picture
does not represent the full state space (e.g., states containing terms like
$a_i\,\sigma^i \otimes \mathbbm{1}$ or $b_i\,\mathbbm{1} \otimes \sigma^i$ are not
contained on the picture), we have to use Lemma~\ref{lemsepablenearest} to check if the
estimated state $\tilde{\rho}$ is indeed the nearest separable state $\rho_0$.

\subsubsection{Region I}

Let us consider first the entangled states located in the triangle region that includes
the Bell state $| \phi^+ \rangle\,$, i.e. Region I in Fig.~\ref{figqubit}. For $\beta =
0$ the states represent the isotropic state \eqref{rhodefiso}, thus all results have to
agree in this case with Eqs.~\eqref{quditisoresult1}--\eqref{BequalsD} for $d=2$. An
entangled state in Region I is characterized by points, i.e. by the parameter pair
($\alpha$,$\beta$), constrained by
\begin{equation}
    \alpha \;\leq\; \beta + 1, \quad\; \alpha \;\leq\; - \beta +1, \quad\; \alpha \;>\;
    \frac{1}{3}\beta + \frac{1}{3} \,.
\end{equation}
The point in the separable region of Fig.~\ref{figqubit} that is nearest (in the
Euclidean sense) to the point ($\alpha$,$\beta$) is given by ($\frac{1}{3} +
\frac{1}{3}\beta$,$\beta$), which corresponds to the state
\begin{equation} \label{rhosep-qubitpauli}
    \tilde{\rho}_\beta \;=\; \frac{1}{4} \left( \mathbbm{1} \,+\,
    \frac{1+\beta}{3} \,
    \left( \sigma_1 \otimes \sigma_1 \,-\, \sigma_2 \otimes \sigma_2
    \right) \,+\, \frac{1-2\beta}{3} \, \sigma_3 \otimes \sigma_3 \right) \,.
\end{equation}
For the difference of nearest--separable and entangled state we obtain
\begin{equation} \label{rhominusqubit}
    \tilde{\rho}_\beta \,-\, \rho^{\rm{ent}}_{\alpha, \beta} \;=\; \frac{1}{4}
    \left( \frac{1+\beta}{3} \,-\, \alpha \right) \Sigma \,,
\end{equation}
where $\Sigma$ is defined by
\begin{equation}
    \Sigma \;:=\; \sigma_1 \otimes \sigma_1 \,-\, \sigma_2 \otimes \sigma_2
    \,+\, \sigma_3 \otimes \sigma_3 \,.
\end{equation}
Using the norm $\| \Sigma \| = 2 \sqrt{3}$ we gain the HS distance
\begin{equation} \label{hsqubit}
    \| \tilde{\rho}_\beta - \rho_{\alpha,\beta}^{\rm{ent}} \| \;=\;
    \frac{\sqrt{3}}{2} \left( \alpha - \frac{1}{3} - \frac{1}{3}
    \beta \right) \,.
\end{equation}
To check whether the state $\tilde{\rho}_\beta$ coincides with the nearest separable
state $\rho_{0; \beta}$ in the sense of the HS measure of entanglement \eqref{defhs}
(which has to take into account the \emph{whole} set of separable states), we have to
test ---according to Lemma~\ref{lemsepablenearest}--- whether the operator
\begin{equation}
    \tilde{C} \;=\; \frac{ \tilde{\rho}_\beta - \rho^{\rm{ent}}_{\alpha,\beta} \,-\,
    \langle \tilde{\rho}_\beta, \tilde{\rho}_\beta -
    \rho^{\rm{ent}}_{\alpha,\beta} \rangle \,\mathbbm{1} }{ \| \tilde{\rho}_\beta -
    \rho_{\alpha,\beta}^{\rm{ent}} \| }
\end{equation}
is an entanglement witness. Remember that any entanglement witness $A$ that detects the
entanglement of a state $\rho^{\rm{ent}}$ has to satisfy the inequalities
\eqref{defentwit}.

We calculate
\begin{equation}
    \langle \tilde{\rho}_\beta , \tilde{\rho}_\beta - \rho^{\rm{ent}}_{\alpha,\beta}
    \rangle \;=\; \textnormal{Tr} \,\tilde{\rho}_\beta ( \tilde{\rho}_\beta -
    \rho^{\rm{ent}}_{\alpha,\beta} ) \;=\; -\frac{1}{4} \left( \alpha - \frac{1}{3} -
    \frac{1}{3} \beta \right)
\end{equation}
and use Eqs.~\eqref{rhominusqubit} and \eqref{hsqubit} to determine the operator
$\tilde{C}$ for the considered case,
\begin{equation} \label{concretectilde}
    \tilde{C} \;=\; \frac{1}{2 \sqrt{3}} \,\left( \mathbbm{1} \,-\, \Sigma \right) \,.
\end{equation}
Then we find
\begin{equation} \label{Centinequal}
    \langle \rho^{\rm{ent}}_{\alpha,\beta}, \tilde{C} \rangle \;=\;
    - \frac{\sqrt{3}}{2} \left( \alpha - \frac{1}{3} - \frac{1}{3}
    \beta \right) \;<\; 0 \,,
\end{equation}
since the entangled states in the considered Region I satisfy the constraint $\alpha >
\frac{1}{3}{\beta} + \frac{1}{3}\,$. Thus the first condition of inequalities
\eqref{defentwit} is fulfilled.

Actually, condition \eqref{Centinequal} is just a consistency check for the correct
calculation of operator $\tilde{C}$ since by construction of $\tilde{C}$ we always have $
\langle \rho^{\rm{ent}}, \tilde{C} \rangle \;=\; - \| \tilde{\rho} - \rho^{\rm{ent}} \|
\;<\; 0 \,$. Thus more important is the test of the second condition of inequalities
\eqref{defentwit} and in order to do it we need the following lemma:
\begin{lem} \label{lemqubit}
    For any Hermitian operator $C$ that is of the form
    \begin{equation} \label{lemqubitc}
        C \;=\; a\, \left(\mathbbm{1} \,+\, c_1 \left( \sigma_x \otimes \sigma_x \,-\,
            \sigma_y \otimes \sigma_y \right) \,+\, c_2\, \sigma_z \otimes
            \sigma_z\right) \qquad a \in \mathbbm{R^+}, \quad -1 \leq c_i \leq 1
    \end{equation}
    the expectation value for all separable states is positive,
    \begin{equation}
        \langle \rho , C \rangle \,\geq\, 0 \qquad \forall \rho \in S \,.
    \end{equation}
\end{lem}
\emph{Proof.} Any separable state $\rho$ is a convex combination of product states and
thus a separable two--qubit state can be written as the Bloch vector (see
Refs.~\cite{bertlmann02, bertlmann05})
\begin{eqnarray} \label{seppauli}
    & \rho \;=\; \sum_k p_k \, \frac{1}{4} \left(
        \mathbbm{1} \otimes \mathbbm{1}
        \,+\, \sum_i n_i^k\,\sigma^i \otimes \mathbbm{1} \,+\, \sum_j m_j^k\,
        \mathbbm{1} \otimes \sigma^j
        \,+\, \sum_{i,j} n_i^k m_j^k \,\sigma^i \otimes \sigma^j \right)\,, & \nonumber\\
    & \textrm{with}\quad n_i^k, m_i^k \in \mathbbm{R}\,, \; \left| \vec{n}^k \right| \leq
        1\,, \left| \vec{m}^k \right| \leq 1 \,, \ \ p_k \geq 0, \, \sum_k p_k = 1
        \,. &
\end{eqnarray}
Performing the trace we obtain
\begin{equation}
    \langle \rho , C \rangle \;=\; \textnormal{Tr} \, \rho \, C \;=\; \sum_k p_k
    \,a \left( 1 \,+\, c_1 \left( n_x^k m_x^k \,-\, n_y^k m_y^k \right) \,+\,
    c_2\, n_z^k m_z^k \right) \,.
\end{equation}
We have
\begin{equation}
    \left| c_1 \left( n_x^k m_x^k \,-\, n_y^k m_y^k \right) \,+\, c_2 \, n_z^k
    m_z^k \right| \;\leq\; |n_x^k| |m_x^k| + |n_y^k| |m_y^k| + |n_z^k|
    |m_z^k| \;\leq\; 1 \,,
\end{equation}
and therefore
\begin{equation}
    \langle \rho , C \rangle \;\geq\; 0 \qquad \forall \rho \in S \,. \quad \Box
\end{equation}
\\

Since the operator $\tilde{C}$ \eqref{concretectilde} is of the form \eqref{lemqubitc} we
can use Lemma~\ref{lemqubit} to verify
\begin{equation}
    \langle \rho , \tilde{C} \rangle \;\geq\; 0 \qquad \forall \rho \in S \,.
\end{equation}
Therefore $\tilde{C}$ \eqref{concretectilde} is indeed an entanglement witness and
$\tilde{\rho}_\beta$ is the nearest separable state $\tilde{\rho}_\beta = \rho_{0; \,
\beta}$ for the entangled states $\rho_{\alpha,\beta}^{\rm{ent}}$ in Region I.\\

Finally, we find for the HS measure of the states in Region I
\begin{equation} \label{hsqubitreal}
    D(\rho_{\alpha, \beta}^{\rm{ent}}) \;=\; \| \rho_{0; \, \beta} \,-\,
    \rho_{\alpha,\beta}^{\rm{ent}} \| \;=\;
    \frac{\sqrt{3}}{2} \left( \alpha - \frac{1}{3} - \frac{1}{3} \beta \right) \,.
\end{equation}

\subsubsection{Region II}

It remains to determine the HS measure for the entangled states
$\rho_{\alpha,\beta}^{\rm{ent}}$ located in the triangle region that includes the Bell
state $| \phi^- \rangle\,$, i.e. Region II in Fig.~\ref{figqubit}. Here the entangled
states are characterized by points $(\alpha, \beta)\,$, where the parameters are
constrained by
\begin{equation}
    \alpha \;\leq\; \beta + 1, \quad\; \alpha \;\geq\; \frac{1}{3} \beta -
    \frac{1}{3}, \quad\; \alpha \;<\; - \beta - 1 \,.
\end{equation}
The states in the separable region of Fig.~\ref{figqubit} that are nearest to the
entangled states $(\alpha, \beta)$ in Region II are called $\tilde{\rho}_{\alpha, \beta}$
and characterized by the points
\begin{equation}
\left(
    \begin{array}{c}
    \tilde{\alpha} \\
    \tilde{\beta}
    \end{array}
\right) \;=\; \left(
    \begin{array}{c}
    1/3 \,\left ( -1 + 2 \alpha - \beta \right) \\
    1/3 \,\left ( -2 - 2 \alpha + \beta \right)
    \end{array}
\right) \,.
\end{equation}
The necessary quantities for calculating the operator $\tilde{C}$ are the following
\begin{equation}
    \tilde{\rho}_{\alpha, \beta} - \rho_{\alpha, \beta}^{\rm{ent}} \;=\;
    - \frac{1}{12} \left( \alpha + 1 + \beta \right) \left( \sigma_x
    \otimes \sigma_x - \sigma_y \otimes \sigma_y - \sigma_z \otimes
    \sigma_z \right) \,,
\end{equation}
\begin{equation}
    \| \tilde{\rho}_{\alpha, \beta} - \rho_{\alpha, \beta}^{\rm{ent}} \| \;=\;
    \frac{1}{2\sqrt{3}} \left( - \alpha - 1 - \beta \right) \,,
\end{equation}
\begin{equation}
    \langle \tilde{\rho}_{\alpha, \beta} , \tilde{\rho}_{\alpha, \beta}^{\rm{ent}} -
    \rho_{\alpha,\beta} \rangle \;=\; \frac{1}{12} \left( \alpha + 1 + \beta \right) \,,
\end{equation}
so that $\tilde{C}$ is expressed by
\begin{equation} \label{concretectilde2}
    \tilde{C} \;=\; \frac{1}{2\sqrt{3}} \left( \mathbbm{1} \,+\, \sigma_1
    \otimes \sigma_1 \,-\, \sigma_2 \otimes \sigma_2 \,-\, \sigma_3 \otimes
    \sigma_3 \right) \,.
\end{equation}
To test $\tilde{C}$ for being an entanglement witness we need to check the first
condition of inequalities \eqref{defentwit}; we get
\begin{equation}
    \langle \rho_{\alpha, \beta}^{\rm{ent}} , \tilde{C} \rangle \;=\;
    \frac{1}{2\sqrt{3}} \left( \alpha + 1 + \beta \right) \;<\; 0
\end{equation}
as expected. Since operator $\tilde{C}$ \eqref{concretectilde2} is of the form
\eqref{lemqubitc} we apply Lemma~\ref{lemqubit} and obtain for the separable states
\begin{equation}
    \langle \rho , \tilde{C} \rangle \;\geq\; 0 \qquad \forall \rho \in S \,.
\end{equation}
Therefore also in Region II operator $\tilde{C}$ \eqref{concretectilde2} is indeed an
entanglement witness and $\tilde{\rho}_{\alpha,\beta}$ is the nearest separable state
$\tilde{\rho}_{\alpha,\beta} = \rho_{0; \, \alpha \beta}$ for the entangled states
$\rho_{\alpha,\beta}^{\rm{ent}}$.\\

For the HS measure of the states in Region II we find
\begin{equation}
    D(\rho_{\alpha,\beta}^{\rm{ent}}) \;=\; \| \rho_{0; \, \alpha, \beta} \,-\,
    \rho_{\alpha,\beta}^{\rm{ent}} \| \;=\; \frac{1}{2\sqrt{3}}
    \left( - \alpha - 1 - \beta \right) \,.
\end{equation}

\subsection{Two--parameter entangled states --- qutrits}\label{secqutrits}

The procedure of determining the geometry of separable and entangled states discussed in
Sec.~\ref{secqubits} can be generalized to higher dimensions, e.g. for two--qutrit
states. Let us first notice how to generalize the concept of a maximally entangled Bell
basis to higher dimensions. A basis of maximally entangled two--qudit states can be
attained by starting with a maximally entangled qudit state $| \phi_0 \rangle$ and
constructing the other $d^2-1$ states in the following way:
\begin{equation}
    | \phi_i \rangle \;=\; \tilde{U}_i \otimes \mathbbm{1} \,| \phi_0
    \rangle \qquad i=1,2, \ldots ,d^2-1 \,,
\end{equation}
where $\{ \tilde{U}_i \}$ represents an orthogonal matrix basis \eqref{orthogon} of
\emph{unitary} matrices and $\tilde{U}_0$ usually denotes the unity matrix $\mathbbm{1}$
(see Refs. \cite{vollbrecht00, werner01}).

A reasonable choice to start with is the maximally entangled state $| \phi_+^d \rangle$
\eqref{defmaxent} and using the WOB (see Sec.~\ref{secwob}) which is an orthogonal basis
of unitary matrices. Such a construction has been proposed in Ref.~\cite{narnhofer06}.
Then we set up the following $d^2$ projectors onto the maximally entangled states -- the
Bell states:
\begin{equation}
    P_{nk} \;:=\; (U_{nk} \otimes \mathbbm{1}) \,| \phi_+^d \rangle \langle \phi_+^d
    |\, (U_{nk}^\dag \otimes \mathbbm{1}) \qquad n,k = 0,1, \ldots ,d-1
    \,.
\end{equation}
We can express the Bell projectors as Bloch vectors by using the Bloch vector form
\eqref{maxentwob2} of $P_{00} := | \phi_+^d \rangle \langle \phi_+^d |$ and the relations
(indices have to be taken $mod \, d$) \cite{narnhofer06}
\begin{eqnarray}
    U_{nm}^\dag & \;=\; & e^{\frac{2 \pi i}{d}nm} \,U_{-n \, -k} \,, \\
    U_{nm}U_{lk} & \;=\; & e^{\frac{2 \pi i}{d}ml} \,U_{n+l \, m+k} \,.
\end{eqnarray}
It provides for the Bell projector the Bloch form
\begin{equation} \label{projbloch}
    P_{nk} \;=\; \frac{1}{d^2} \sum_{m,l=0}^{d-1} e^{\frac{2 \pi
    i}{d}(kl-nm)} \,U_{lm} \otimes U_{-lm} \,.
\end{equation}

In case of qutrits ($d=3$) the 9 Bell projectors \eqref{projbloch} form an 8--dimensional
simplex which is the higher dimensional analogue of a 3--dimensional simplex, the
tetrahedron for qubits, see Fig.~\ref{figtetrahedron}. This 8--dimensional simplex has a
very interesting geometry concerning separability and entanglement (see
Refs.~\cite{baumgartner06, baumgartner07}). Due to its high symmetry inside
---named therefore the \emph{magic simplex} by the authors of
Ref.~\cite{baumgartner06}--- it is enough to consider certain mixtures of Bell states
which form equivalent classes concerning their geometry.

We are interested in the following two--parameter states of two--qutrits as a
generalization of the qubit case, Eq.~\eqref{rhoqubit},
\begin{equation} \label{rhoqutrit}
    \rho_{\alpha, \beta} \;=\; \frac{1- \alpha -\beta}{9} \,\mathbbm{1} \,+\,
    \alpha \,P_{00} \,+\, \beta \frac{1}{2}
    \left( P_{10} + P_{20} \right)\,.
\end{equation}
According to Ref.~\cite{baumgartner06} the Bell states represent points in a discrete
phase space. The indices $n,k$ of the Bell states can be interpreted as ``quantized''
position coordinate and momentum, respectively. The Bell states $P_{00},P_{10}$ and
$P_{20}$ lie on a line in this phase space picture of the maximally entangled states,
they exhibit the same geometry as other lines since each line can be transformed into
another one.

\begin{figure}
  \begin{centering}
    \includegraphics[width=0.7\textwidth]{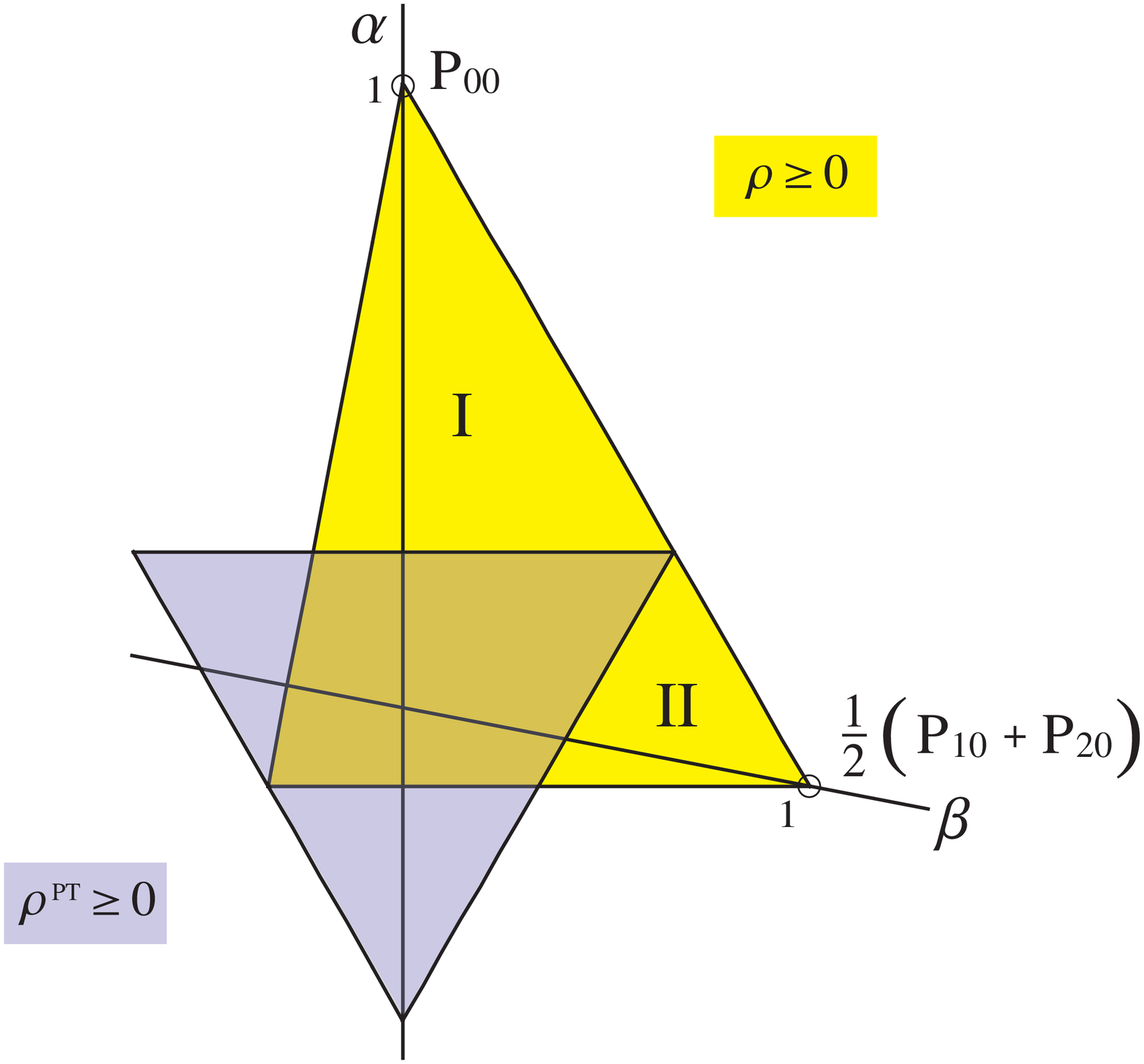}\\
    \caption{Illustration of the two--qutrit states $\rho_{\alpha, \beta}$ \eqref{rhoqutrit}.}
    \label{figqutrit}
  \end{centering}
\end{figure}

Inserting the Bloch vector form of $P_{00},P_{10}$ and $P_{20}$ \eqref{projbloch} we find
the Bloch vector expansion of the two--parameter states \eqref{rhoqutrit}
\begin{equation} \label{rhoqutritweyl}
    \rho_{\alpha, \beta} \;=\; \frac{1}{9} \left( \mathbbm{1} \,+\, \left(\alpha -
    \frac{\beta}{2} \right) U_1 \,+\, \left( \alpha + \beta \right)U_2 \right) \,,
\end{equation}
where we defined
\begin{eqnarray} \label{defu1u2}
    U_1 & \;:=\; & U_{01} \otimes U_{01} + U_{02} \otimes U_{02} + U_{11}
    \otimes U_{-11} + U_{12} \otimes U_{-12} + U_{21} \otimes
    U_{-21}
    + U_{22} \otimes U_{-22} \,, \nonumber\\
    U_2 & \;:=\; & U_{10} \otimes U_{-10} \,+\, U_{20} \otimes U_{-20} \,.
\end{eqnarray}
The constraints for the positivity requirement ($\rho_{\alpha, \beta} \geq 0$) are
\begin{equation}
    \alpha \;\leq\; \frac{7}{2} \beta +1, \quad\; \alpha \;\leq\; -\beta +1, \quad\;
    \alpha \;\geq\; \frac{\beta}{8} - \frac{1}{8} \,,
\end{equation}
and for the PPT
\begin{equation}
    \alpha \;\leq\; - \beta -\frac{1}{2}, \quad\; \alpha \;\geq\; \frac{5}{4}\beta -\frac{1}{2},
    \quad\; \alpha \;\leq\; \frac{\beta}{8} + \frac{1}{4} \,.
\end{equation}
The Euclidean picture representing the HS space geometry of states \eqref{rhoqutrit} is
shown in Fig.~\ref{figqutrit}, where the parameter coordinate axes are non--orthogonal
since in HS space they have to be orthogonal to the boundary lines of the positivity
region, $\alpha = \frac{\beta}{8} - \frac{1}{8}$ and $\alpha = \frac{7}{2} \beta +1$.
It is shown in Ref.~\cite{baumgartner06} that the PPT states $\rho_{\alpha, \beta}$ are
all separable states, so there are no bound entangled PPT states of the form
\eqref{rhoqutrit}. To find the HS measure for the entangled two--parameter 2--qutrit
states we apply the same procedure as in Sec.~\ref{secqubits}: We determine the states
that are the nearest separable ones in the Euclidean sense of Fig.~\ref{figqutrit} and
use Lemma~\ref{lemsepablenearest} to check whether these are indeed the nearest separable
ones with respect to the whole state space (for other approaches see, e.g.,
Refs.~\cite{verstraete02, cao07}).

\subsubsection{Region I}

First we consider Region I in Fig.~\ref{figqutrit}, i.e., the triangle region of
entangled states around the $\alpha$-axis, constrained by the parameter values
\begin{equation}
    \alpha \;\leq\; \frac{7}{2} \beta +1 , \quad\; \alpha \leq -\beta +1 , \quad\;
    \alpha > \frac{\beta}{8} + \frac{1}{4} \,.
\end{equation}
In the Euclidean picture the point that is nearest to point $(\alpha, \beta)$ in this
region is given by $(\frac{1}{4}+\frac{1}{8}\beta, \beta)$, which corresponds to the
separable two--qutrit state
\begin{equation}
    \tilde{\rho}_\beta \;=\; \frac{1}{9} \left( \mathbbm{1} \,+\, \left(
    \frac{1}{4} - \frac{3}{8}\beta \right) U_1 \,+\, \left( \frac{1}{4}
    + \frac{9}{8}\beta \right) U_2 \right) \,,
\end{equation}
with $U_1$ and $U_2$ defined in Eq.~\eqref{defu1u2}.

For the difference of nearest--separable and entangled state we find
\begin{equation} \label{rhominusqutrit}
    \tilde{\rho}_\beta - \rho^{\rm{ent}}_{\alpha, \beta} \;=\;
    \frac{1}{9} \left( \frac{1}{4} + \frac{1}{8}\beta - \alpha \right) U \,,
\end{equation}
where $U = U_1 + U_2$ (and is defined in Eq.~\eqref{defu}), and using for the norm $\| U
\| = d \sqrt{8} = 6 \sqrt{2}$ we gain the HS distance
\begin{equation} \label{hsqutrit}
    \| \tilde{\rho}_\beta - \rho_{\alpha,\beta}^{\rm{ent}} \| \;=\;
    \frac{2\sqrt{2}}{3} \left( \alpha - \frac{1}{4} - \frac{1}{8}
    \beta \right) \,.
\end{equation}
It remains to calculate
\begin{equation}
    \langle \tilde{\rho}_\beta , \tilde{\rho}_\beta - \rho_{\alpha,\beta}
    \rangle \;=\; \textnormal{Tr} \,\tilde{\rho}_\beta ( \tilde{\rho}_\beta -
    \rho_{\alpha,\beta} ) \;=\; -\frac{2}{9} \left( \alpha - \frac{1}{4} + \frac{1}{8}
    \beta \right)
\end{equation}
to set up the operator
\begin{equation} \label{ctildequtrit}
    \tilde{C} \;=\; \frac{ \tilde{\rho}_\beta - \rho^{\rm{ent}}_{\alpha,\beta} \,-\,
    \langle \tilde{\rho}_\beta, \tilde{\rho}_\beta - \rho^{\rm{ent}}_{\alpha,\beta}
    \rangle \,\mathbbm{1} }{ \| \tilde{\rho}_\beta - \rho_{\alpha,\beta}^{\rm{ent}} \| } \;=\;
    \frac{1}{6\sqrt{2}} \,\left( 2\,\mathbbm{1} \,-\, U \right) \,.
\end{equation}
We test now whether it represents an entanglement witness, i.e., whether $\tilde{C}$
\eqref{ctildequtrit} satisfies the inequalities \eqref{defentwit}. As expected we find
\begin{equation}
    \langle \rho^{\rm{ent}}_{\alpha,\beta}, \tilde{C} \rangle \;=\; - \frac{2\sqrt{2}}{3}
    \left( \alpha - \frac{1}{4} - \frac{1}{8} \beta \right) \;<\; 0 \,.
\end{equation}
To check the second condition of inequalities \eqref{defentwit} we set up the following
lemma, similar to Lemma~\ref{lemqubit}:
\begin{lem} \label{lemqutrit}
    For any Hermitian operator $C$ that is of the form
    \begin{equation} \label{lemqutritc}
        C \;=\; a \,(2\,\mathbbm{1} \,+\, c_1 \,U_1 \,+\, c_2 \,U_2) \qquad a \in \mathbbm{R^+},
        \quad -1 \leq c_i \leq 1
    \end{equation}
    the expectation value for all separable states is positive,
    \begin{equation}
        \langle \rho , C \rangle \;\geq\; 0 \qquad \forall \rho \in S \,.
    \end{equation}
\end{lem}
\emph{Proof.} Any separable two--qutrit state can be written as the Bloch vector
\cite{bertlmann02, bertlmann05}
\begin{eqnarray} \label{sepweyl}
    \rho & \;=\; & \sum_k p_k \ \frac{1}{9} \Big( \mathbbm{1} \otimes \mathbbm{1}
        \,+\, \sum_{n,m=0}^{d-1} \sqrt{2}\, n_{nm}^k \,U_{nm} \otimes \mathbbm{1}
        \,+\, \sum_{l,k=0}^{d-1} \sqrt{2}\, m_{lk}^k \,\mathbbm{1} \otimes U_{lk}
        \nonumber\\
    & & \,+\, \sum_{n,m,l,k=0}^{d-1} 2\, n_{nm}^k m_{lk}^k \,U_{nm} \otimes U_{lk} \Big) \,,
        \nonumber\\
    & & \textrm{with} \quad n_{nm}^k, m_{lk}^k \in \mathbbm{C}\,, \; \left| \vec{n}^k \right|
        \leq 1\,, \left| \vec{m}^k \right| \leq 1 \,, \; \ p_k \geq 0, \, \sum_k p_k = 1 \,,
\end{eqnarray}
where we define $\left| \vec{n}^k \right|^2 := \sum_{nm} n_{nm}^*n_{nm}\,$.

Performing the trace we obtain
\begin{equation}
    \langle \rho , C \rangle \;=\; \textnormal{Tr} \,\rho^\dag \, C \;=\;
    \sum_k p_k \left( 2a \left( 1 \,+\, c_1 \sum_{i=1}^6 n_i^k m_i^k \,+\,
    \sum_{j=7}^8 c_2 n_j^k m_j^k \right) \right) \,,
\end{equation}
with
\begin{eqnarray}
    \left( n_1, n_2, n_3, n_4, n_5, n_6, n_7, n_8 \right) & \;=\; & \left(
    n_{01}^*, n_{02}^*, n_{11}^*, n_{12}^*, n_{21}^*, n_{22}^*, n_{10}^*,
    n_{20}^* \right) \mbox{ and} \nonumber\\
    \left( m_1, m_2, m_3, m_4, m_5, m_6, m_7, m_8 \right) & \;=\; & \left(
    m_{01}^*, m_{02}^*, m_{-11}^*, m_{-12}^*, m_{-21}^*, m_{-22}^*, m_{-10}^*,
    m_{-20}^* \right) . \nonumber\\
    & &
\end{eqnarray}
We have
\begin{equation}
    \left| c_1 \sum_{i=1}^6 n_i^k m_i^k \,+\, c_2 \sum_{j=7}^8 n_j^k m_j^k \right|
    \;\leq\; \sum_i |n_i^k| |m_i^k| \;\leq\; 1
\end{equation}
and therefore
\begin{equation}
    \langle \rho , C \rangle \;\geq\; 0 \qquad \forall \rho \in S \,. \quad \Box
\end{equation}

Since the operator $\tilde{C}$ \eqref{ctildequtrit} is of the form \eqref{lemqutritc} we
can use Lemma~\ref{lemqutrit} to verify
\begin{equation}
    \langle \rho , \tilde{C} \rangle \;\geq\; 0 \qquad \forall \rho \in S \,.
\end{equation}
Thus $\tilde{C}$ \eqref{ctildequtrit} is indeed an entanglement witness and
$\tilde{\rho}_\beta$ is the nearest separable state $\tilde{\rho}_\beta = \rho_{0; \,
\beta}$ for the entangled states $\rho_{\alpha,\beta}^{\rm{ent}}$ in Region I.\\

For the HS measure of the entangled two--parameter two--qutrit states \eqref{rhoqutrit}
we find
\begin{equation} \label{hsqutritreal}
    D(\rho_{\alpha, \beta}^{\rm{ent}}) \;=\; \| \rho_{0; \, \beta} -
    \rho_{\alpha,\beta}^{\rm{ent}} \| \;=\; \frac{2\sqrt{2}}{3}
    \left( \alpha - \frac{1}{4} - \frac{1}{8} \beta \right) \,.
\end{equation}

\subsubsection{Region II}

In Region II of Fig.~\ref{figqutrit} the entangled two--parameter two--qutrit states are
constrained by
\begin{equation}\label{param-qutrit-regionII}
    \alpha \;<\; \frac{5}{4}\beta - \frac{1}{2}, \quad\; \alpha \;\geq\;
    \frac{1}{8}\beta - \frac{1}{8}, \quad\; \alpha \;\leq\; -\beta + 1 \,.
\end{equation}
The points that have minimal Euclidean distance to the points $(\alpha, \beta)$ located
in this region are characterized by
\begin{equation}
\left(
    \begin{array}{c}
    \tilde{\alpha} \\
    \tilde{\beta}
    \end{array}
\right) = \left(
    \begin{array}{c}
    1/24 \,\left( -2 + 20\alpha + 5\beta \right) \\
    1/6 \,\left( 2 + 4\alpha + \beta \right)
    \end{array}
\right) \,,
\end{equation}
and correspond to the states $\tilde{\rho}_{\alpha, \beta}$. The quantities needed for
calculating $\tilde{C}$ are
\begin{eqnarray}
    \tilde{\rho}_{\alpha, \beta} \,-\, \rho_{\alpha, \beta}^{\rm{ent}} &\;=\;&
    - \frac{1}{72} \left( 4\alpha + 2 - 5\beta \right) \left( U_1 - U_2 \right) \,,\\
    \| \tilde{\rho}_{\alpha, \beta} - \rho_{\alpha, \beta}^{\rm{ent}} \| &\;=\;&
    \frac{1}{6\sqrt{2}} \left( -4\alpha - 2 + 5\beta \right) \,,\\
    \langle \tilde{\rho}_{\alpha, \beta} , \tilde{\rho}_{\alpha, \beta}^{\rm{ent}} -
    \rho_{\alpha, \beta} \rangle &\;=\;& \frac{1}{36} \left( 4\alpha + 2 - 5\beta \right) \,,
\end{eqnarray}
so that operator $\tilde{C}$ is expressed by
\begin{equation} \label{ctildequtrit2}
    \tilde{C} \;=\; \frac{1}{6\sqrt{2}} \,\left( 2\,\mathbbm{1} \,+\, U_1 \,-\, U_2 \right) \,.
\end{equation}
The check of the first condition \eqref{defentwit} for an entanglement witness gives,
unsurprisingly,
\begin{equation}
    \langle \rho_{\alpha, \beta}^{\rm{ent}} , \tilde{C} \rangle \;=\;
    \frac{1}{6\sqrt{2}} \,\left( 4\alpha + 2 - 5\beta \right) \,<\, 0 \,,
\end{equation}
since $4\alpha < 5\beta -2\,$, Eq.~\eqref{param-qutrit-regionII}. For the second test we
use the fact that operator $\tilde{C}$ \eqref{ctildequtrit2} is of the form
\eqref{lemqutritc} and thus, according to Lemma~\ref{lemqutrit}, we obtain
\begin{equation}
    \langle \rho , \tilde{C} \rangle \;\geq\; 0 \qquad \forall \rho \in S \,.
\end{equation}
Therefore $\tilde{C}$ \eqref{ctildequtrit2} is indeed an entanglement witness and the
states $\tilde{\rho}_{\alpha, \beta}$ are the nearest separable ones
$\tilde{\rho}_{\alpha, \beta} = \rho_{0; \, \alpha, \beta}$ to the entangled
two--parameter states \eqref{rhoqutrit} of Region II.

Finally, for the HS measure of these states we obtain
\begin{equation}
    D(\rho_{\alpha,\beta}^{\rm{ent}}) \;=\; \| \rho_{0; \, \alpha, \beta} \,-\,
    \rho_{\alpha,\beta}^{\rm{ent}} \| \;=\; \frac{1}{6\sqrt{2}} \,
    \left( -4\alpha - 2 + 5\beta \right) \,.\\
\end{equation}

\vspace{0.3cm}

\section{Conclusion}
\label{conclusion}

In this paper we present three different matrix bases which are quite useful to decompose
density matrices for higher dimensional qudits. These are the generalized Gell-Mann
matrix basis GGB, the polarization operator basis POB, and the Weyl operator basis WOB.
Each decomposition we identify with a vector, the so-called Bloch vector.

Considering just one--particle states we observe the following features:

The GGB is easy to construct, the matrices correspond to the standard SU(N) generators ($N=d$),
but in general (in $d$ dimensions) it is rather unpractical to work with the diagonal GGM
\eqref{ggmd} due to its more complicated definition. On the other hand, the Bloch vector itself
has real components, which is advantageous, they can be expressed as expectation values of
measurable quantities. For example, in $3$ dimensions the Gell-Mann matrices are Hermitian and the
Bloch vector components can be expressed by expectation values of spin 1 operators. The POB is
also easy to set up, all you need to know are the Clebsch--Gordan coefficients which you find
tabulated in the literature. However, the Bloch vector contains complex components. The Weyl
operators for the WOB are also simple to construct, they are non--Hermitian but unitary operators.
The Bloch vector itself turns out very simple, however, with complex components. Let us note that
in $2$ dimensions all bases are equivalent since they correspond to Pauli matrices or linear
combinations thereof.

In case of two--qudits we have studied the isotropic states explicitly and find the following
characteristics:

In the GGB the Bloch vector \eqref{isoggb} with expression \eqref{Bloch-Lambda} is more
complicated to construct, in particular the diagonal part $B$ \eqref{ggb-term-B} (see
Appendix \ref{termB}). In the POB the Bloch vector \eqref{isopob} with expression
\eqref{Bloch-T} can be easily set up by the knowledge of the Clebsch--Gordon coefficient
sum rule \eqref{CGsumrule} and in the WOB the Bloch vector \eqref{isowob} with definition
\eqref{defu} is actually most easily to construct.

The Hilbert--Schmidt measure of entanglement has been calculated explicitly for the isotropic
two--qudit states and we want to emphasize its connection to the maximal violation of a
generalized Bell inequality (Theorem~\ref{BNT-theorem}), an inequality for the entanglement
witness.

We demonstrate the geometry of separability and entanglement in case of qubits by choosing
so-called two--parameter states, Eq.~\eqref{rhoqubit}, i.e., planes in the tetrahedron formed by
the Bell states (see Fig.~\ref{figtetrahedron}). These states reflect already the underlying
geometry of the Hilbert Space and they are chosen with regard to the description of qutrit states,
a generalization into higher dimensions. To a given entangled state we determine the nearest
separable state, calculate the corresponding entanglement witness and the Hilbert--Schmidt measure
in the relevant Regions I and II (see Fig.~\ref{figqubit}).

In case of qutrits it is quite illustrative to demonstrate the geometry of separability and
entanglement in terms of two--parameter states \eqref{rhoqutrit}. These states set up a plane in
the 8--dimensional simplex formed by the $9$ Bell states and are easy to construct within the WOB.
Due to the high symmetry of the \emph{magic} simplex we may restrict ourselves to a certain
mixture of Bell states, Eq.~\eqref{rhoqutrit}, which exhibits the same geometry as other lines.
Within the WOB it is quite easy to find the Bloch vector form \eqref{rhoqutritweyl} of the
two--parameter states. It is straightforward to find for a given entangled state in the relevant
Regions I and II (see Fig.~\ref{figqutrit}) the nearest separable state and the corresponding
entanglement witness. The easy calculation of the Hilbert--Schmidt measure of entanglement is a
great advantage in this case and its result of high interest since it is quite difficult to
calculate other entanglement measures for higher dimensional states, like the entanglement of
formation.

It turns out that the Weyl operator basis is optimal for all our calculations, the reason is that
entanglement ---the maximally entangled Bell states--- is in fact easily constructed by unitary
operators \`a la Weyl.

\begin{acknowledgments}

We would like to thank Heide Narnhofer, Beatrix Hiesmayr, Alexander
Ableitinger and Marcus Huber for helpful discussion and comments.
This research has been supported by the ``F140-N
Forschungsstipendium'' of the University of Vienna.

\end{acknowledgments}

\appendix

\section{}

\subsection{Proof of Orthogonality of GGB}\label{orthogonality}

We want to proof condition \eqref{orthogon} for the GGB which consists of the $d^2-1$ GGM
\eqref{ggms}, \eqref{ggma}, \eqref{ggmd} and the $d \times d$ unity $\mathbbm{1}$. Since
all GGM are Hermitian (thus $\textnormal{Tr} A_i^{\dag} A_j = \textnormal{Tr} A_i A_j =
\textnormal{Tr} A_j A_i$) it suffices to proof the following conditions:
\begin{eqnarray}
    \textnormal{Tr} \,\Lambda^{jk}_s \Lambda^{mn}_s & \;=\; & 2 \,\delta^{jm}
        \delta^{kn} \label{proofprop5}\\
    \textnormal{Tr} \,\Lambda^{jk}_a \Lambda^{mn}_a & \;=\; & 2 \,\delta^{jm}
        \delta^{kn} \label{proofprop6}\\
    \textnormal{Tr} \,\Lambda^{l} \Lambda^{m} & \;=\; & 2 \,\delta^{lm}
        \label{proofprop7} \\
    \textnormal{Tr} \,\Lambda^{jk}_a \Lambda^{mn}_s & \;=\; & 0 \label{proofprop8} \\
    \textnormal{Tr} \,\Lambda^{jk}_s \Lambda^{m} & \;=\; & 0 \label{proofprop9} \\
    \textnormal{Tr} \,\Lambda^{jk}_a \Lambda^{m} & \;=\; & 0 \,. \label{proofprop10}
\end{eqnarray}

\emph{Proof of condition \eqref{proofprop5}.} Inserting definition \eqref{ggms} we have
\begin{eqnarray} \label{proofss}
    \textnormal{Tr} \,\Lambda^{jk}_s \Lambda^{mn}_s & \;=\; &
        \sum_{l=1}^{d} \langle l | \left( |j \rangle \langle k | \,+\, |
        k  \rangle \langle j | \right) \left( |m \rangle \langle n | \,+\, |
        n  \rangle \langle m | \right) | l \rangle \nonumber\\
    & \;=\; & \sum_l \left(
    \langle l | j \rangle \langle k | m \rangle \langle n | l \rangle \,+\,
    \langle l | j \rangle \langle k | n \rangle \langle m | l \rangle \,+\,
    \langle l | k \rangle \langle j | m \rangle \langle n | l \rangle \,+\,
    \langle l | k \rangle \langle j | n \rangle \langle m | l
    \rangle \right) \nonumber\\
    & \;=\; & \delta^{jn} \delta^{km} \,+\, \delta^{jm} \delta^{kn} \,+\,
        \delta^{kn} \delta^{jm} \,+\, \delta^{km} \delta^{jn} \nonumber\\
    & \;=\; & 2 \,\delta^{jm} \delta^{kn} \,,
\end{eqnarray}
where we used in the last step that $\delta^{jn} \delta^{km} = 0$ since we have $j<k$
\emph{and} $m<n$.\\

\emph{Proof of condition \eqref{proofprop6}.} This case is equivalent to the one before
apart from changed signs that do not matter
\begin{eqnarray} \label{proofaa}
    \textnormal{Tr} \,\Lambda^{jk}_a \Lambda^{mn}_a & \;=\; &
        -\,\delta^{jn} \delta^{km} \,+\, \delta^{jm} \delta^{kn} \,+\,
        \delta^{kn} \delta^{jm} \,-\, \delta^{km} \delta^{jn} \nonumber\\
    & \;=\; & 2 \,\delta^{jm} \delta^{kn} \,.
\end{eqnarray}

\emph{Proof of condition \eqref{proofprop7}.} Using definition \eqref{ggmd} and denoting
\begin{equation} \label{defcdiagggm}
    C_l = \sqrt{\frac{2}{l(l+1)}} \,,
\end{equation}
where $l \leq m$ without loss of generality, we get
\begin{eqnarray}
    \textnormal{Tr} \,\Lambda^l \Lambda^m & \;=\; & C_l C_m \sum_{p=1}^d
        \Big( \sum_{k=1}^l \sum_{n=1}^m \langle p | k \rangle
        \langle k | n \rangle \langle n | p \rangle \,+\, l m \langle p
        | l+1 \rangle \langle l+1 | m+1 \rangle \langle m+1 | p
        \rangle \nonumber\\
    & & - \,m \sum_{k=1}^l \langle p | k \rangle \langle k | m+1
        \rangle \langle m+1 | p \rangle \,-\, l \sum_{n=1}^m \langle p |
        l+1 \rangle \langle l+1 | n \rangle \langle n | p \rangle
        \Big)
        \nonumber\\
    & \;=\; & C_l C_m \left( l \,+\, l m \, \delta^{lm} \,-\, m \sum_{k=1}^{l}
    \delta ^{k(m+1)} \,-\, l \sum_{n=1}^{m} \delta ^{n(l+1)} \right) \,.
\end{eqnarray}
Using the fact that $\delta^{k (m+1)} = 0$ for $m \geq k$ and
\begin{equation}
    l \sum_{n=1}^m \delta^{n (l+1)} \;=\; \begin{cases}
        0 & \text{if} \ l = m \\
        l & \text{if} \ l < m
        \end{cases}
\end{equation}
we obtain
\begin{equation}
    \textnormal{Tr} \,\Lambda^l \Lambda^m \;=\; (C_l)^2 \, l (l+1) \,
    \delta^{lm} \;=\; 2 \,\delta^{lm} \,.
\end{equation}

\emph{Proof of condition \eqref{proofprop8}.} Analogously to the proofs \eqref{proofss}
and \eqref{proofaa} we find
\begin{equation}
    \textnormal{Tr} \,\Lambda^{jk}_a \Lambda^{mn}_s \;=\; i \left(
    - \,\delta^{jn} \delta^{km} \,+\, \delta^{jm} \delta^{kn} \,-\, \delta^{jm}
    \delta^{kn} \,+\, \delta^{jn} \delta^{km} \right) \;=\; 0 \,.
\end{equation}

\emph{Proof of condition \eqref{proofprop9}.} Inserting definitions \eqref{ggms} and
\eqref{ggmd} gives
\begin{eqnarray}
    \text{Tr} \,\Lambda^{jk}_s \Lambda^m & \;=\; & C_m \sum_{p=1}^d \Big(
        - m \langle p | k \rangle \langle j | m+1 \rangle \langle m+1 | p
        \rangle \,-\, m \langle p | j \rangle \langle k | m+1 \rangle
        \langle m+1 | p \rangle \nonumber\\
    & & + \sum_{n=1}^{m} \langle p | j \rangle \langle k | n \rangle
        \langle n | p \rangle \,+\, \sum_{n=1}^m \langle p | k \rangle
        \langle j | n \rangle \langle n | p \rangle \Big)
        \nonumber\\
    & \;=\; & - \,2m \,\delta^{j (m+1)} \delta^{k (m+1)} \,+\, 2 \sum_{l=1}^m
    \delta^{kl} \delta^{jl} \nonumber\\
    & \;=\; & 0 \,,
\end{eqnarray}
since per definition we have $j < k\,$.\\

\emph{Proof of condition \eqref{proofprop10}.} This proof is equivalent to the previous
one since constant factors in front of the terms do not matter.

\subsection{Calculation of term B in GGB}\label{termB}

To obtain the Bloch vector notation of term $B$ \eqref{isoggmB} we insert the standard
matrix expansion \eqref{smggb} for the case $j = k$. We split the tensor products in the
following way
\begin{equation}
    B \;=\; \frac{1}{d} \left( B_1 \,+\, B_2 \,+\, B_3 \,+\, B_4 \,+\,
    \frac{1}{d} \,\mathbbm{1} \otimes \mathbbm{1} \right) \,,
\end{equation}
where the terms $B_1, \ldots ,B_4$ are introduced by (note that $\Lambda^0 = 0$)
\begin{eqnarray}
    B_1 & \;=\; & \sum_{j=1}^d \left( \frac{j-1}{2j} \Lambda^{j-1} \otimes
        \Lambda^{j-1} \,+\, \sum_{n(=l)=0}^{d-j-1} \frac{1}{2(j+n)(j+n+1)} \Lambda^{j+n} \otimes
        \Lambda^{j+n} \right) \label{isoggbb1} \\
    B_2 & \;=\; & \sum_{j=1}^d \Bigg( - \sum_{l=0}^{d-j-1} \sqrt{\frac{j-1}
        {4j(j+l)(j+l+1)}} \Lambda^{j-1} \otimes \Lambda^{j+l}
        \nonumber\\
        & & \qquad \ \, - \sum_{n=0}^{d-j-1} \sqrt{\frac{j-1}
        {4j(j+n)(j+n+1)}} \Lambda^{j+n} \otimes \Lambda^{j-1} \nonumber\\
        & & \qquad \ \, + \sum_{n \neq l,\, n,l=0}^{d-j-1} \frac{1}{2
        \sqrt{(j+n)(j+n+1)(j +
        l)(j+l+1)}} \Lambda^{j+n} \otimes \Lambda^{j+l} \Bigg)
        \label{isoggbb2} \\
    B_3 & \;=\; & \frac{1}{d} \sum_{j=1}^d \left( - \sqrt{\frac{j-1}{2j}}
        \Lambda^{j-1} \otimes \mathbbm{1} \,+\, \sum_{n=0}^{d-j-1}
        \frac{1}{\sqrt{2(j+n)(j+n+1)}} \Lambda^{j+n} \otimes
        \mathbbm{1} \right) \label{isoggb3} \\
    B_4 & \;=\; & \frac{1}{d} \sum_{j=1}^d \left( - \sqrt{\frac{j-1}{2j}}
        \mathbbm{1} \otimes \Lambda^{j-1} \,+\, \sum_{l=0}^{d-j-1}
        \frac{1}{\sqrt{2(j+l)(j+l+1)}} \mathbbm{1} \otimes
        \Lambda^{j+l} \right) \label{isoggb4} \,.
\end{eqnarray}
Only the first term $B_1$ \eqref{isoggbb1} gives a contribution
\begin{equation}
    B_1 \;=\; \sum_{m=1}^{d-1} \left( \frac{m}{2(m+1)} \,+\,
    \frac{m}{2m(m+1)} \right) \Lambda^m \otimes \Lambda^m \;=\; \frac{1}{2}
    \sum_{m=1}^{d-1} \Lambda^m \otimes \Lambda^m \,,
\end{equation}
whereas the remaining terms vanish:
\begin{eqnarray}
    B_2 & \;=\; & \sum_{m < p,\, m,p =1}^{d-1} \left( -
        \sqrt{\frac{m}{4(m+1)p(p+1)}} \,+\, \frac{m}{\sqrt{4m(m+1)p(p+1)}}
        \right) \Lambda^m \otimes \Lambda^p \nonumber\\
    & & + \sum_{m > p,\, m,p =1}^{d-1} \left( -
        \sqrt{\frac{p}{4(p+1)m(m+1)}} \,+\, \frac{p}{\sqrt{4p(p+1)m(m+1)}}
        \right) \Lambda^m \otimes \Lambda^p \nonumber\\
    & \;=\; & \left( \sum_{m<p} \frac{-m+m}{2 \sqrt{m(m+1)p(p+1)}} \,+\,
        \sum_{m>p} \frac{-p+p}{2 \sqrt{m(m+1)p(p+1)}} \right) \Lambda^m
        \otimes \Lambda^p \nonumber\\
    & \;=\; & 0 \,,
\end{eqnarray}
and in quite the same manner
\begin{eqnarray}
    B_3 & \;=\; & \frac{1}{d} \sum_{m=1}^{d-1}
        \frac{-m+m}{\sqrt{2m(m+1)}} \ \Lambda^m \otimes \mathbbm{1} = 0 \,, \nonumber\\
    B_4 & \;=\; & \frac{1}{d} \sum_{p=1}^{d-1}
        \frac{-p+p}{\sqrt{2p(p+1)}} \ \mathbbm{1} \otimes \Lambda^p = 0 \,.
\end{eqnarray}
Thus we find the following Bloch vector of $B$ \eqref{isoggmB}
\begin{equation}
    B \;=\; \frac{1}{2d}
    \sum_{m=1}^{d-1} \Lambda^m \otimes \Lambda^m \,+\, \frac{1}{d^2} \,\mathbbm{1} \otimes
    \mathbbm{1} \,.
\end{equation}

\subsection{Proof of Orthonormality of WOB} \label{secproofon}

For proofs relevant in the WOB we often need the equivalence
\begin{equation} \label{complexsumrule}
    \sum_{n=0}^{d-1} e^{\frac{2 \pi i}{d} \,nx} \;=\; \begin{cases}
        d & \text{if } x=0 \\
        0 & \text{if } x\neq 0
    \end{cases}, \quad x \in \mathbbm{Z} \,.
\end{equation}
So we use Eq.~\eqref{complexsumrule} to proof the orthonormality of the Weyl operators
\eqref{defwo}
\begin{eqnarray}
    \text{Tr} \,U_{nm}^{\dag} U_{lj} & \;=\; & \sum_{p=0}^{d-1} \sum_{k,
        \tilde{k} = 0}^{d-1} e^{\frac{2 \pi i}{d}\,(\tilde{k}l-kn)} \,\langle p |
        (k+m) \,\textrm{mod}\,d \rangle \langle k | \tilde{k} \rangle \langle
        (\tilde{k} + j ) \,\textrm{mod}\,d |p \rangle \nonumber\\
    & \;=\; & \sum_{p=0}^{d-1} \sum_{k,
        \tilde{k} = 0}^{d-1} e^{\frac{2 \pi i}{d}\,(\tilde{k}l-kn)} \,\langle p |
        (k+m) \,\textrm{mod}\,d \rangle  \langle
        (\tilde{k} + j ) \,\textrm{mod}\,d |p \rangle \,\delta_{k \tilde{k}} \nonumber\\
    & \;=\; & \sum_{k=0}^{d-1} e^{\frac{2 \pi i}{d}\,k (l-n)} \,\delta_{mj}
        \nonumber\\
    & \;=\; & d \,\delta_{nl} \,\delta_{mj} \,.
\end{eqnarray}

\subsection{Expansion into WOB}\label{BellWOB}

Formula \eqref{maxentwob2} for the Bell state in terms of WOB we derive in the following
way. We express the standard matrices by the WOB \eqref{smwob}, rewrite the indices and
separate the nonvanishing terms
\begin{eqnarray} \label{maxentwob1}
  \left| \phi_+^d \right\rangle \left\langle \phi_+^d \right|
    & \;=\; & \frac{1}{d} \sum_{j,k =1}^{d} |j \rangle \langle k |
        \otimes |j \rangle \langle k| \nonumber\\
    & \;=\; & \frac{1}{d^3} \sum_{j,k=0}^{d-1} \sum_{l,l'=0}^{d-1} e^{-\frac{2 \pi i}{d}\,j
        (l+l')} U_{l (k-j) mod \, d} \otimes U_{l' (k-j) mod \, d}
        \nonumber\\
    & \;=\; & \frac{1}{d^3} \sum_{m,k=0}^{d-1} \sum_{l,l'=0}^{d-1} e^{-\frac{2 \pi i}{d}\,
        (k-m)(l+l')} U_{l m} \otimes U_{l' m} \nonumber\\
    & \;=\; & \frac{1}{d^2} \left( \sum_m U_{0m} \otimes U_{0m} \,+\,
        \sum_m \, \sum_{l,l';\,l+l'=d} U_{lm} \otimes U_{l'm} \right) \nonumber\\
    & & +\; \frac{1}{d^3} \sum_m \, \sum_{
        l,l';\,l,l' \neq 0,0;\,l+l' \neq d} \left( \sum_k e^{-\frac{2 \pi i}{d}\,(k-m) (l+l')}
       \right) U_{lm} \otimes U_{l'm} \,.
\end{eqnarray}
The last term in Eq.~\eqref{maxentwob1} vanishes due to relation \eqref{complexsumrule}.
Identifying $U_{00} = \mathbbm{1}$ and using the notation with negative values of the
index $l$, which have to be considered as $mod \ d\,$, we gain the formula
\begin{equation}
    \left| \phi_+^d \right\rangle \left\langle \phi_+^d \right| \;=\;
    \frac{1}{d^2} \,\mathbbm{1} \otimes \mathbbm{1} \,+\, \frac{1}{d^2}
    \,\sum_{l,m = 0}^{d-1} U_{lm} \otimes U_{-lm} \,, \qquad (l,m) \neq (0,0) \,.
\end{equation}

\bibliography{references}

\end{document}